\documentclass[aps,prd,twocolumn,superscriptaddress,groupedaddress]{revtex4}  
\usepackage{graphicx}  
\usepackage{dcolumn}   
\usepackage{bm}        
\usepackage{amssymb}   
\usepackage[usenames]{color}

\begin{document}



\title{Measurements of the cosmological parameters $\Omega_m$, $\Omega_k$, 
$\Omega_\textrm{\small{de}}(a)$, $H_0$, and $\sum m_\nu$}
\author{B.~Hoeneisen} \affiliation{Universidad San Francisco de Quito, Quito, Ecuador}
\date{November 19, 2018}

\begin{abstract}
\noindent
From Baryon Acoustic Oscillation measurements with
Sloan Digital Sky Survey SDSS DR14 galaxies,
and the acoustic horizon angle $\theta_*$
measured by the Planck Collaboration, we obtain
$\Omega_m = 0.2724 \pm 0.0047$, and
$h + 0.020 \cdot \sum{m_\nu} = 0.7038 \pm 0.0060$,
assuming flat space and a cosmological constant.
We combine this result with the
2018 Planck ``TT,TE,EE$+$lowE$+$lensing" analysis,
and update a study of $\sum m_\nu$ with
new direct measurements of $\sigma_8$, and obtain
$\sum m_\nu = 0.27 \pm 0.08$ eV
assuming three nearly degenerate neutrino eigenstates.
Measurements are consistent with $\Omega_k = 0$,
and $\Omega_\textrm{\small{de}}(a) = \Omega_\Lambda$ 
constant.
\end{abstract}

\maketitle

\section{Introduction and summary}

From a study of Baryon Acoustic Oscillations (BAO) with
Sloan Digital Sky Survey (SDSS) data release DR13 galaxies and the
``sound horizon" angle $\theta_\textrm{\tiny{MC}}$ measured by the
Planck Collaboration we obtained 
$\Omega_m = 0.281 \pm 0.003$ 
assuming flat space and a cosmological constant \cite{BH_ijaa}. 
At the time, the 2016 Review of Particle Physics
quoted $\Omega_m = 0.308 \pm 0.012$ \cite{PDG2016}.
The new 2018 Planck ``TT,TE,EE$+$lowE$+$lensing"
measurement \cite{Planck} obtains $\Omega_m = 0.3153 \pm 0.0073$,
while the ``TT,TE,EE$+$lowE$+$lensing$+$BAO" 
measurement obtains $\Omega_m = 0.3111 \pm 0.0056$ \cite{Planck}.
Due to the growing tension between these measurements, we 
decided to repeat the BAO analysis in Reference \cite{BH_ijaa}, this time
with SDSS DR14 galaxies. 

The main difficulty with
the BAO measurements is to distinguish the BAO
signal from the cosmological and statistical fluctuations.
The aim of the present
analysis is to be very conservative by choosing
large bins in redshift $z$ to obtain a larger
significance of the BAO signal than in \cite{BH_ijaa}.
As a result, the present analysis is based on 6 
independent BAO measurements, compared to 18 in \cite{BH_ijaa}.

We assume flat space, i.e. $\Omega_k = 0$, and constant dark
energy density, i.e. $\Omega_\textrm{\small{de}}(a) = \Omega_\Lambda$, except in Tables
\ref{galaxy_fits}, \ref{galaxy_theta_fits}, and 
\ref{galaxy_Lya_theta_fits} that include more general cases.
We assume three neutrino flavors with eigenstates with nearly the
same mass, so $\sum m_\nu \approx 3 m_\nu$.
We adopt the notation of the Particle Data Group 2018 \cite{PDG2018}.
All uncertainties have 68\% confidence.

The analysis presented in this article obtains
$\Omega_m = 0.2724 \pm 0.0047$ so the tension has increased
further. We present full details
of all fits to the galaxy-galaxy distance histograms 
of the present measurement so
that the reader may cross-check each step of the analysis.
Calibrating the BAO standard ruler we obtain
$h + 0.020 \cdot \sum{m_\nu} = 0.7038 \pm 0.0060$.

Combining the direct measurement
$\Omega_m = 0.2724 \pm 0.0047$ with the 
2018 Planck ``TT,TE,EE$+$lowE$+$lensing"
analysis obtains
$\Omega_m = 0.2853 \pm 0.0040$ and
$h = 0.6990 \pm 0.0030$, at the cost of an
increase of the Planck $\chi^2_\textrm{\tiny{P}}$ from
12956.78 to 12968.64.

Finally, we update the measurement of $\sum m_\nu$
of Reference \cite{BH_ijaa_3} with the data of
this Planck$+ \Omega_m$ combination, and two
new direct measurements of $\sigma_8$, and obtain
$\sum m_\nu = 0.27 \pm 0.08$ eV.
This result is sensitive to the accuracy of the
direct measurements of $\sigma_8$.

\section{Measurement of $\Omega_m$ with BAO as an \textit{uncalibrated} standard ruler}

\begin{figure}
\begin{center}
\scalebox{0.465} {\includegraphics{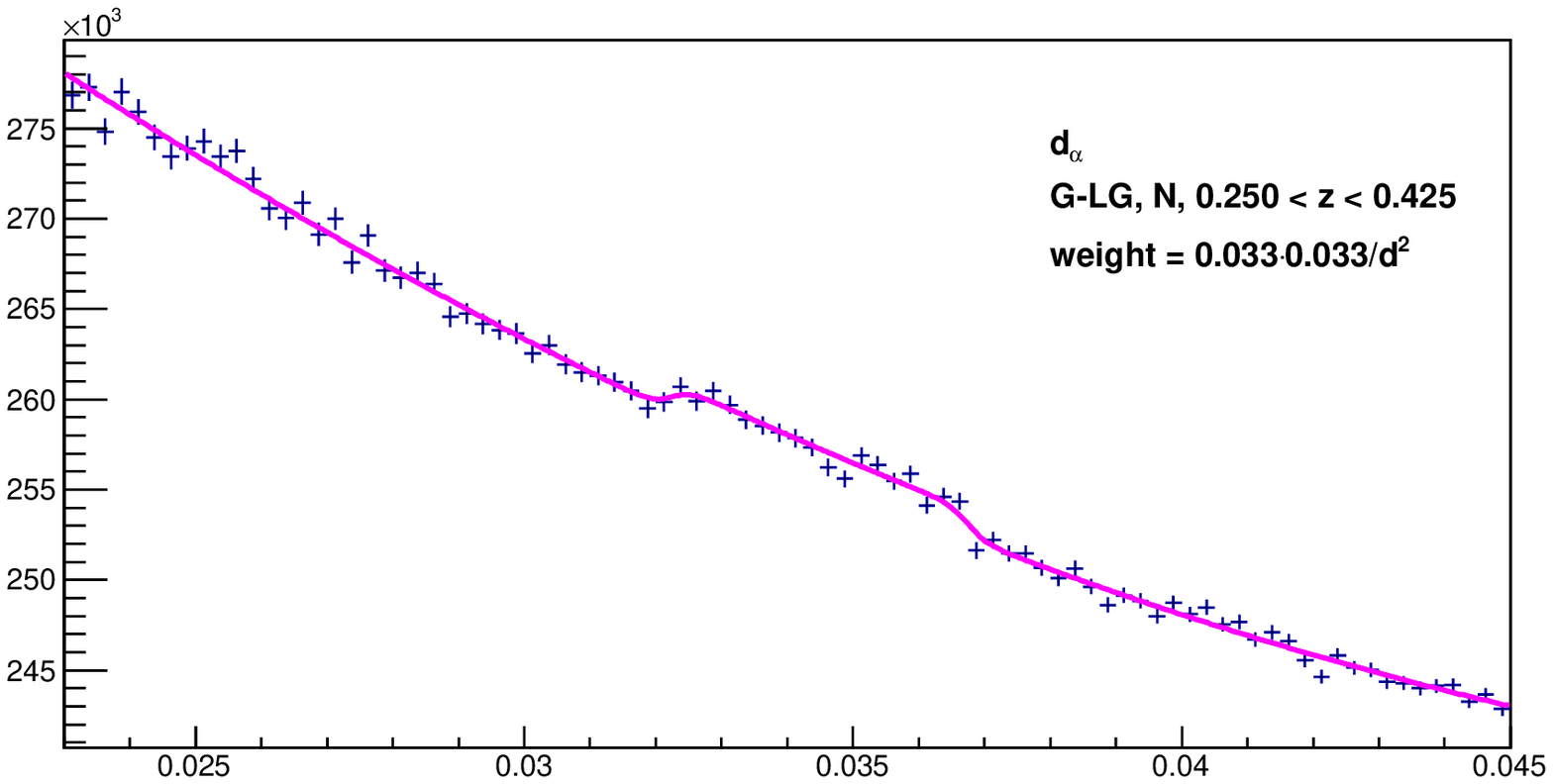}}
\scalebox{0.465} {\includegraphics{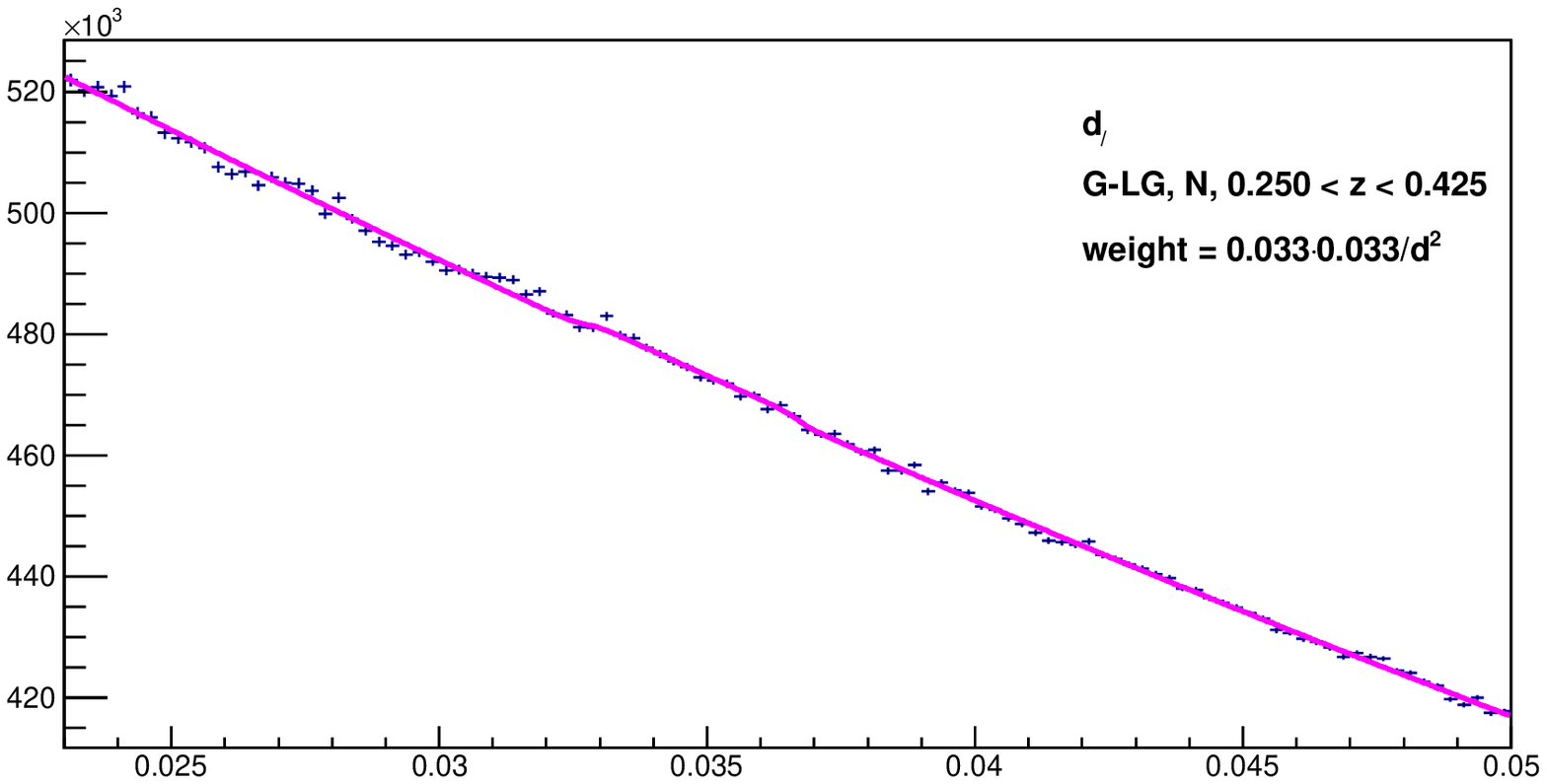}}
\scalebox{0.465} {\includegraphics{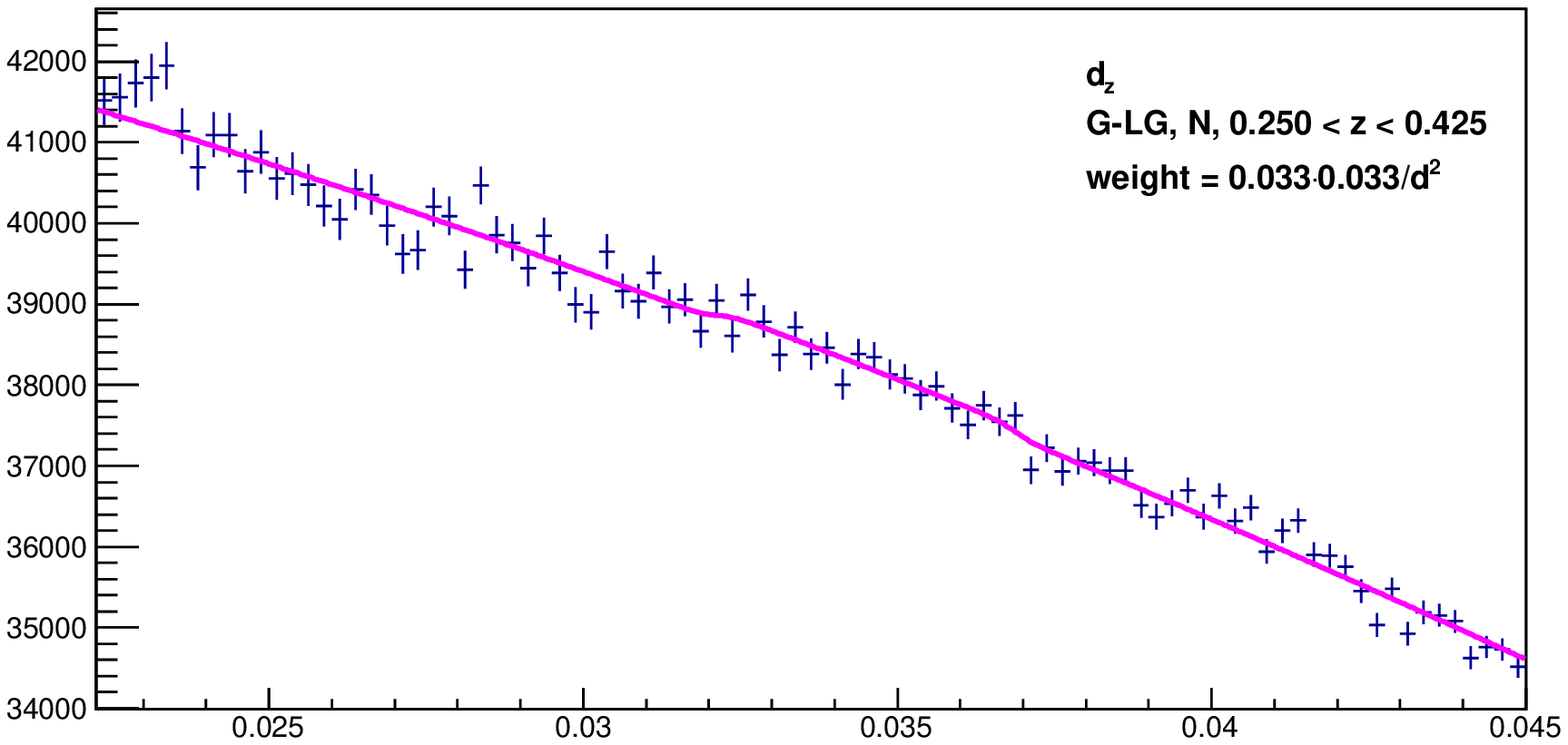}}
\caption{
Fits to histograms of G-LG distances $d$ that obtain
$\hat{d}_\alpha$, $\hat{d}_/$, or $\hat{d}_z$ at $z = 0.34$.
See Tables \ref{d_meas} and \ref{selected} for details.
}
\label{fig_250_425}
\end{center}
\end{figure}

\begin{figure}
\begin{center}
\scalebox{0.465} {\includegraphics{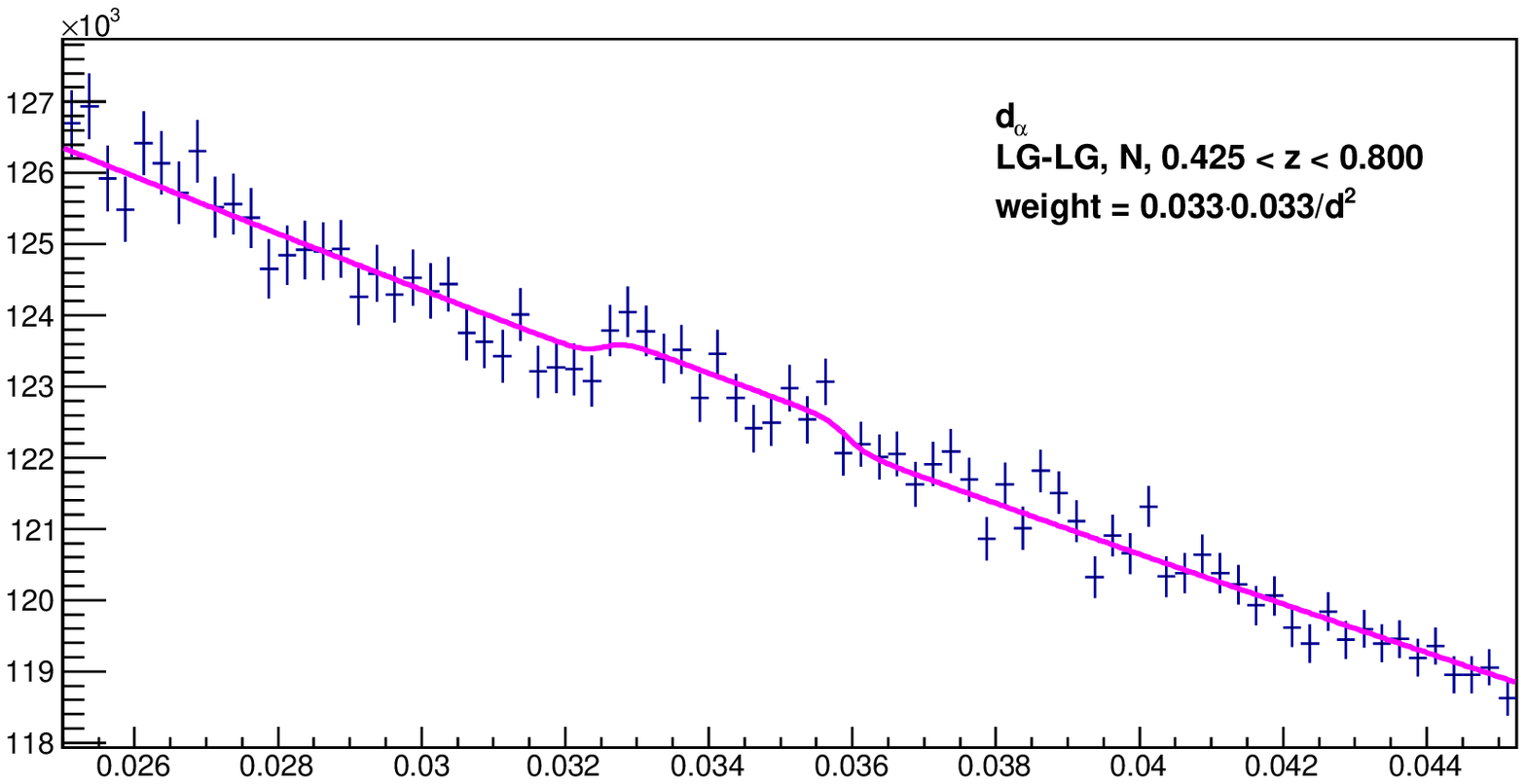}}
\scalebox{0.465} {\includegraphics{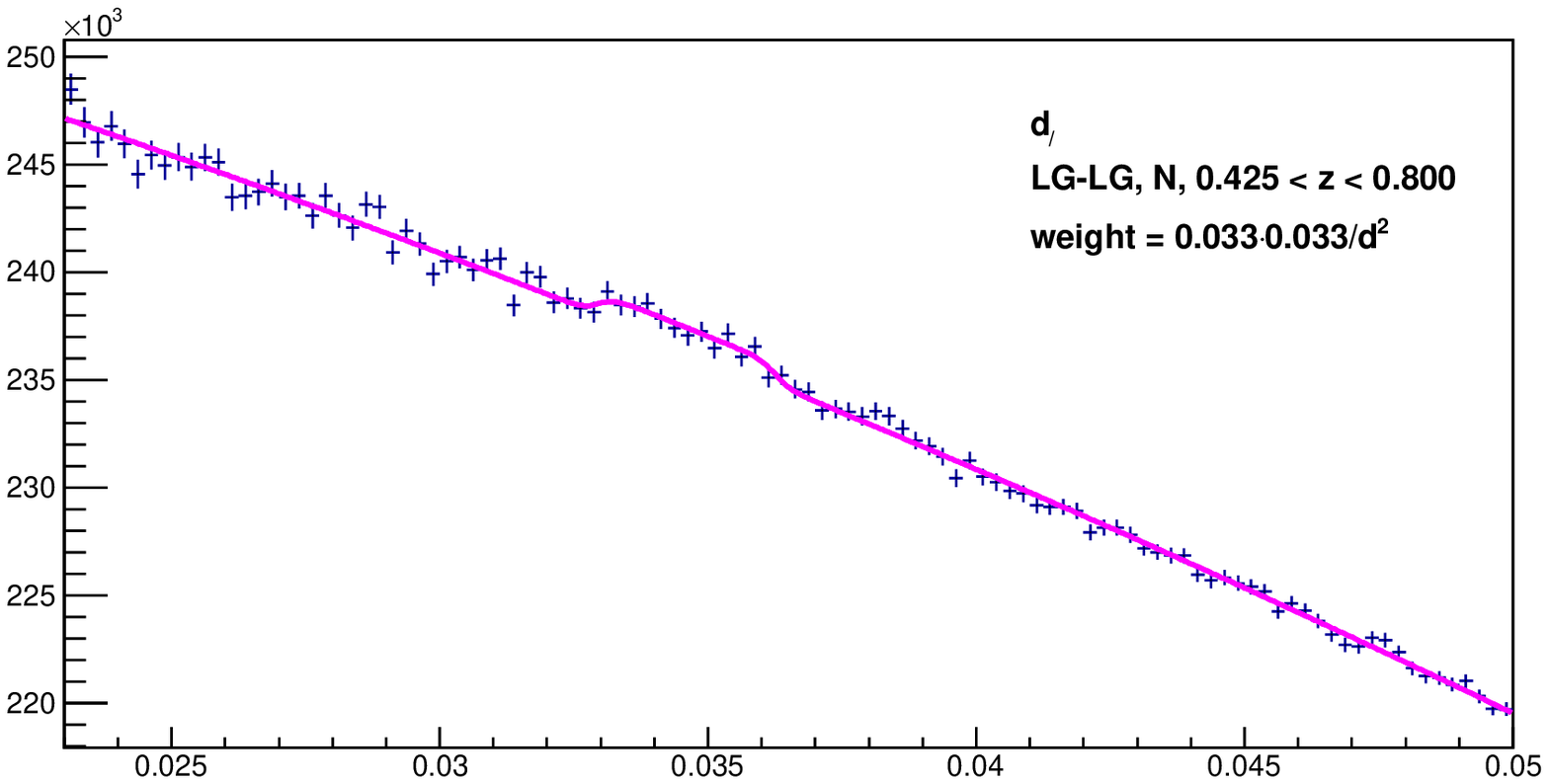}}
\scalebox{0.465} {\includegraphics{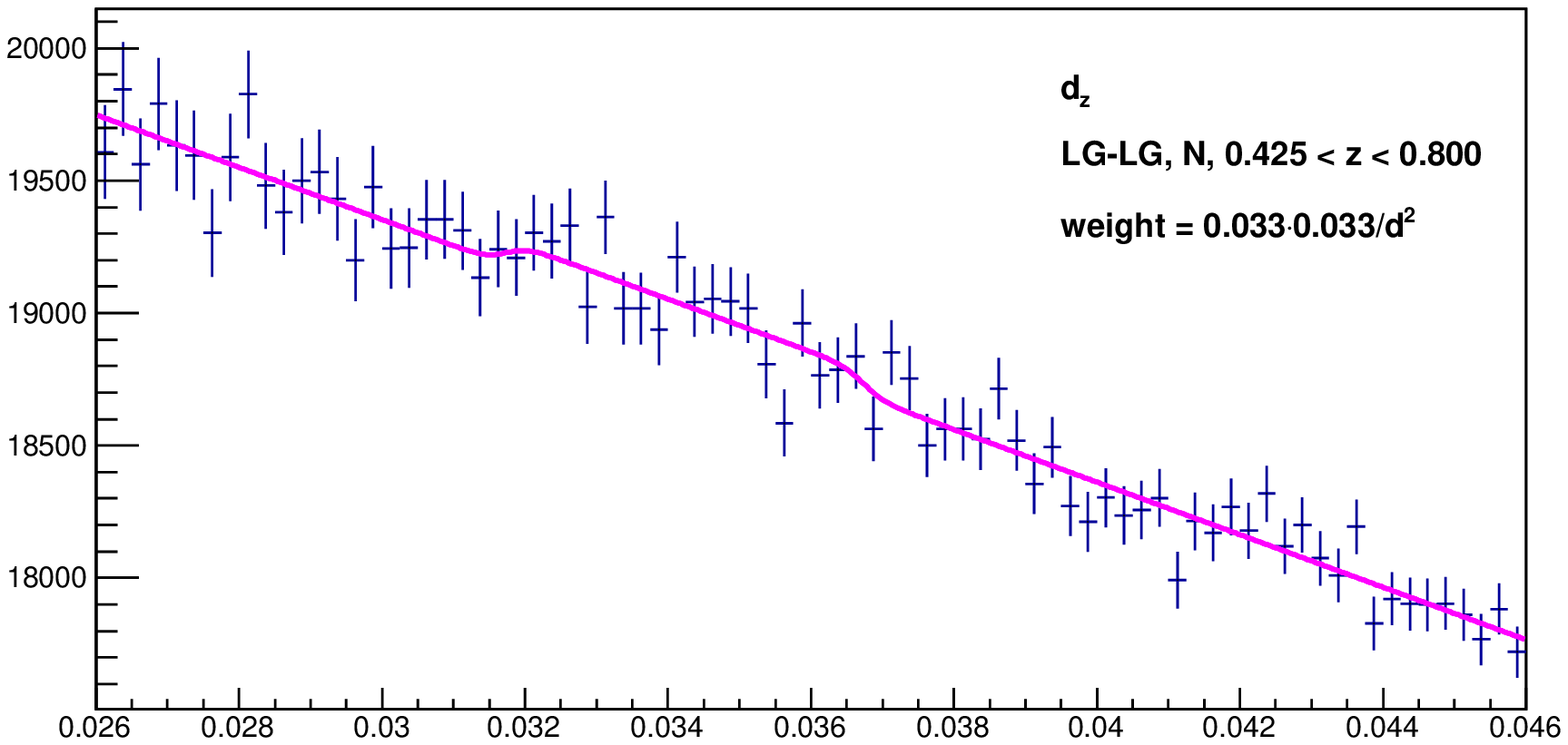}}
\caption{
Fits to histograms of LG-LG distances $d$ that obtain
$\hat{d}_\alpha$, $\hat{d}_/$, or $\hat{d}_z$ at $z = 0.56$.
See Tables \ref{d_meas} and \ref{selected} for details.
}
\label{fig_425_800}
\end{center}
\end{figure}

We measure the comoving galaxy-galaxy correlation distance $d_\textrm{\small{drag}}$,
in units of $c/H_0$, with galaxies in the Sloan Digital Sky Survey SDSS DR14 
publicly released catalog \cite{SDSS_DR14, BOSS},
with the method described in Reference \cite{BH_ijaa}. 
Briefly, from the angle $\alpha$ between two galaxies as seen by the observer, and their
red-shifts $z_1$ and $z_2$, we calculate their distance $d$, in units of $c/H_0$,
assuming a reference cosmology \cite{BH_ijaa}. 
At this ``uncalibrated" stage
in the analysis, the unit of distance $c/H_0$ is neither known nor needed.
The adimensional distance $d$ has a component $d_\alpha$
transverse to the line of sight, and a component $d_z$ along the line of sight,
given by Equation (3) of \cite{BH_ijaa}.
We fill three histograms of $d$ according to the orientation of the galaxy
pairs with respect to the line of sight, i.e. $d_z/d_\alpha < 1/3$, $d_\alpha/d_z < 1/3$,
and remaining pairs. Fitting these histograms we obtain excesses centered at
$\hat{d}_\alpha$, $\hat{d}_z$, and $\hat{d}_/$ respectively.
Examples are shown in Figures \ref{fig_250_425} and \ref{fig_425_800}.
From each BAO observable $\hat{d}_\alpha$, $\hat{d}_/$, or $\hat{d}_z$	
we recover $d_\textrm{\small{drag}}$ for any given cosmology with Equations (5), (6),
or (7) of Ref. \cite{BH_ijaa}. Requiring that $d_\textrm{\small{drag}}$
be independent of red shift $z$ and orientation we obtain
the space curvature $\Omega_k$, the dark energy density $\Omega_\textrm{\small{de}}(a)$
as a function of the expansion parameter $a = 1/(1 + z)$, and the 
matter density $\Omega_m = 1 - \Omega_\textrm{\small{de}}(1) - \Omega_k - \Omega_r$.
Full details can be found in \cite{BH_ijaa}.

The challenge with these BAO measurements is to distinguish the BAO signal
from the cosmological and statistical fluctuations of the background. 
Our strategy is three-fold: (i) redundancy of measurements with
different cosmological fluctuations, (ii) pattern recognition of the
BAO signal, and (iii) requiring all three fits for 
$\hat{d}_\alpha$, $\hat{d}_/$, and $\hat{d}_z$ to converge,
and that the consistency relation 
$Q = \hat{d}_/ / (\hat{d}_\alpha^{0.57} \hat{d}_z^{0.43}) = 1$
\cite{BH_ijaa} be satisfied within $\pm 3\%$. 

Regarding redundancy, we repeat the fits for the northern (N) and
southern (S) galactic caps; we repeat the measurements for galaxy-galaxy (G-G)
distances, galaxy-large galaxy (G-LG) distances, LG-LG distances, and
galaxy-cluster (G-C) distances; and
we fill histograms of $d$ with weights $0.033^2/d^2$ or $0.033^2 F_i F_j / d^2$,
where $F_i$ and $F_j$ are absolute luminosities; see \cite{BH_ijaa} for details.
In the present analysis we have off-set the bins of redshift $z$ with
respect to Reference \cite{BH_ijaa} to obtain different background
fluctuations.

Now consider pattern recognition.
Figures \ref{fig_250_425} and \ref{fig_425_800} show that the BAO signal
is approximately constant from $\approx 0.032$ to $\approx 0.037$, corresponding
to $\approx 137$ Mpc to $\approx 158$ Mpc. This characteristic shape of the BAO
signal can be understood qualitatively
with reference to Figure (1) of \cite{Eisenstein}:
the radial mass profile of
an initial point like adiabatic excess results, well after recombination, in
peaks at radii 17 Mpc and $r_\textrm{\small{drag}} \approx 148$ Mpc,
so we can expect the BAO signal to extend from approximately 
$148-17$ Mpc to $148 + 17$ Mpc, with 
$r_\textrm{\small{drag}}$ at the mid-point. From galaxy simulations
described in \cite{BH_ijaa_3}, 
the smearing of $r_\textrm{\small{drag}}$ due to
galaxy peculiar motions has a standard deviation approximately 7.6 Mpc at $z = 0.5$,
and 8.5 Mpc at $z = 0.3$. So the observed BAO signal has 
an unexpected ``step-up-step-down" shape, and is narrower than implied
by the simulation in reference \cite{Eisenstein}.

The selections of galaxies are as in \cite{BH_ijaa} with the added requirements for SDSS DR14
galaxies that they be ``\textsf{sciencePrimary}" and ``\textsf{bossPrimary}", and have
a smaller redshift uncertainty \textsf{zErr}$ < 0.00025$.

The fitting function has 6 free parameters, corresponding to a second
degree polynomial for the background, and a ``smooth step-up-step-down"
function (described in \cite{BH_ijaa}) with a center $\hat{d}$, a
half-width $\Delta$, and an amplitude $A$ relative to the background.
Each fit used for the final measurements is required to have 
a significance $A / \sigma_A > 2$
(in the analysis of \cite{BH_ijaa} this requirement was $A / \sigma_A > 1$,
which allows more bins of $z$).

Successful triplets of fits are presented in Table \ref{d_meas}.
Note the redundancy of measurements with $0.250 < z < 0.425$ and $0.425 < z < 800$.
The independent triplets of fits selected for further analysis, are indicated
with a ``$*$", and are shown in Figures \ref{fig_250_425} and
\ref{fig_425_800}, with further details presented in Table \ref{selected}.
We note that each measurement of $\hat{d}_\alpha$, $\hat{d}_/$, or $\hat{d}_z$
in Table \ref{d_meas}, together with the sound horizon angle 
$\theta_*$ obtained 
by the Planck experiment \cite{Planck}, is
a sensitive measurement of $\Omega_m$ as shown in Table \ref{d_calc}.

The peculiar motion corrections were studied with the galaxy generator
described in \cite{BH_ijaa_3, BH_galaxies}. Results of these simulations are shown in
Table \ref{Dxpec}, for G-G distances, for two cases: 
``correct $P(k)$" and ``correct $P_\textrm{gal}(k)$".
The ``correct $P(k)$" simulations have the predicted linear power spectrum
of density fluctuations $P(k)$ of the $\Lambda$CDM model (Eq. (8.1.42) of \cite{Weinberg}),
while the ``correct $P_\textrm{gal}(k)$" simulations have a steeper $P(k)$
input so that the generated galaxy power spectrum $P_\textrm{gal}(k)$ matches observations,
see Figure (15) of \cite{BH_ijaa_3}. (The need for the steeper $P(k)$
is currently not understood.) All of these G-G corrections, and also the
corrections for LG-LG and G-C, are in 
agreement, to within a factor 2, 
with the corrections applied in \cite{BH_ijaa} that where taken from
a study in \cite{Seo}. In summary, in the present analysis we apply the same
peculiar motion corrections as in \cite{BH_ijaa}, i.e. 
we multiply the measured BAO distances
$\hat{d}_\alpha$, $\hat{d}_/$, and $\hat{d}_z$,
by correction factors $f_\alpha$, $f_/$, and $f_z$, respectively, where
\begin{eqnarray}
f_\alpha - 1 & = & 0.00320 \cdot a^{1.35}, \nonumber \\
f_/ - 1 & = & 0.00350 \cdot a^{1.35}, \nonumber \\
f_z - 1 & = & 0.00381 \cdot a^{1.35}.
\label{correction}
\end{eqnarray} 
We take half of these corrections as a systematic uncertainty.
The effect of these corrections
is relatively small as shown in Table \ref{galaxy_fits} below.

\begin{table*}
\caption{\label{d_meas}
Measured BAO distances $\hat{d}_\alpha$,
$\hat{d}_/$, and $\hat{d}_z$, in units of $c/H_0$,
with $z_c = 3.79$ (see \cite{BH_ijaa}) from SDSS DR14 galaxies with
right ascension $110^0$ to $270^0$, and declination $-5^0$ to $70^0$,
in the northern (N) and/or southern (S) galactic caps.
Uncertainties are statistical from the fits to the BAO signal.
No corrections have been applied. The independent measurements with a ``$*$" are 
selected for further analysis. 
The corresponding fits are presented in Figures \ref{fig_250_425} and \ref{fig_425_800}, and
details are presented in Table \ref{selected}.
For comparison, 
measurements with a ``\&" correspond to SDSS DR13 data with the galaxy selections of \cite{BH_ijaa}.
}
\begin{ruledtabular}
\begin{tabular}{c|ccrrl|ccc|c}
$z$ & $z_\textrm{min}$ & $z_\textrm{max}$ & Galaxies & Centers & Type &
$100 \hat{d}_\alpha$ & $100 \hat{d}_/$ & $100 \hat{d}_z$ & Q \\
\hline
0.53 & 0.425 & 0.725 & 614724 & 614724 & G-G, N$+$S & $3.488 \pm 0.015$ & $3.504 \pm 0.019$ & $3.466 \pm 0.032$ & 1.007 \\
0.53 & 0.425 & 0.725 & 614724 & 13960 & G-C, N$+$S & $3.381 \pm 0.030$ & $3.401 \pm 0.033$ & $3.395 \pm 0.035$ & 1.004 \\
0.53 & 0.475 & 0.575 & 180696 & 53519 & G-LG, N & $3.424 \pm 0.015$ & $3.314 \pm 0.018$ & $3.242 \pm 0.018$ & 0.991 \\
0.53 & 0.475 & 0.575 & 53519 & 53519 & LG-LG, N & $3.451 \pm 0.030$ & $3.447 \pm 0.059$ & $3.351 \pm 0.022$ & 1.012 \\
0.53 & 0.475 & 0.575 & 180696 & 5045 & G-C, N & $3.427 \pm 0.031$ & $3.331 \pm 0.030$ & $3.316 \pm 0.033$ & 0.986 \\
\hline
0.56 & 0.425 & 0.800 & 230841 & 230841 & G-G, S & $3.441 \pm 0.027$ & $3.422 \pm 0.017$ & $3.497 \pm 0.040$ & 0.988 \\
0.56 & 0.425 & 0.800 & 355737 & 120499 & G-LG, N & $3.425 \pm 0.015$ & $3.465 \pm 0.016$ & $3.351 \pm 0.025$ & 1.021 \\
$*$0.56 & 0.425 & 0.800 & 120499 & 120499 & LG-LG, N & $3.424 \pm 0.021$ & $3.461 \pm 0.018$ & $3.424 \pm 0.039$ & 1.011 \\
\&0.56 & 0.425 & 0.800 & 143778 & 143778 & LG-LG, N & $3.424 \pm 0.014$ & $3.478 \pm 0.015$ & $3.451 \pm 0.026$ & 1.012 \\
0.56 & 0.425 & 0.800 & 586578 & 13206 & G-C, N$+$S & $3.453 \pm 0.038$ & $3.365 \pm 0.044$ & $3.354 \pm 0.028$ & 0.987 \\
\hline
0.52 & 0.425 & 0.575 & 236693 & 236693 & G-G, N & $3.437 \pm 0.031$ & $3.423 \pm 0.026$ & $3.432 \pm 0.025$ & 0.997 \\
0.52 & 0.425 & 0.575 & 236693 & 72297 & G-LG, N & $3.416 \pm 0.017$ & $3.441 \pm 0.012$ & $3.385 \pm 0.018$ & 1.011 \\
0.52 & 0.425 & 0.575 & 72297 & 72297 & LG-LG, N & $3.456 \pm 0.033$ & $3.447 \pm 0.022$ & $3.392 \pm 0.060$ & 1.006 \\
0.48 & 0.425 & 0.525 & 151938 & 4143 & G-C, N & $3.424 \pm 0.051$ & $3.383 \pm 0.026$ & $3.343 \pm 0.062$ & 0.998 \\
0.36 & 0.250 & 0.450 & 114597 & 114597 & G-G, N & $3.456 \pm 0.018$ & $3.386 \pm 0.015$ & $3.318 \pm 0.056$ & 0.997 \\
0.36 & 0.250 & 0.450 & 114597 & 65130 & G-LG, N & $3.455 \pm 0.010$ & $3.358 \pm 0.015$ & $3.293 \pm 0.032$ & 0.992 \\
0.36 & 0.250 & 0.450 & 65130 & 65130 & LG-LG, N & $3.462 \pm 0.016$ & $3.352 \pm 0.025$ & $3.307 \pm 0.039$ & 0.988 \\
\hline
0.34 & 0.250 & 0.425 & 92321 & 92321 & G-G, N & $3.439 \pm 0.013$ & $3.473 \pm 0.015$ & $3.423 \pm 0.076$ & 1.012 \\
0.34 & 0.250 & 0.425 & 149849 & 149849 & G-G, N$+$S & $3.437 \pm 0.014$ & $3.367 \pm 0.013$ & $3.444 \pm 0.042$ & 0.979 \\
$*$0.34 & 0.250 & 0.425 & 92321 & 55980 & G-LG, N & $3.449 \pm 0.008$ & $3.471 \pm 0.013$ & $3.450 \pm 0.034$ & 1.006 \\
\&0.34 & 0.250 & 0.425 & 133729 & 94873 & G-LG, N & $3.431 \pm 0.011$ & $3.469 \pm 0.014$ & $3.383 \pm 0.024$ & 1.017 \\
0.34 & 0.250 & 0.425 & 55980 & 55980 & LG-LG, N & $3.467 \pm 0.019$ & $3.477 \pm 0.015$ & $3.459 \pm 0.045$ & 1.004 \\
\end{tabular}
\end{ruledtabular}
\end{table*}

\begin{table}
\caption{\label{selected}
Details of the fits selected for the final analysis
(indicated by a ``$*$" in Table \ref{d_meas}).
Note that the significance of the fitted signal amplitudes
(relative to the background)
range from 2.1 to 9.8 standard deviations.
}
\begin{ruledtabular}
\begin{tabular}{lccc}
Observable & $z$ & Relative amplitude $A$ & Half-width $\Delta$ \\
\hline
$\hat{d}_\alpha$   & 0.56 & $0.00290 \pm 0.00100$ & $0.00169 \pm 0.00022$ \\ 
$\hat{d}_/$        & 0.56 & $0.00422 \pm 0.00069$ & $0.00164 \pm 0.00020$ \\ 
$\hat{d}_z$        & 0.56 & $0.00505 \pm 0.00226$ & $0.00250 \pm 0.00041$ \\ 
$\hat{d}_\alpha$   & 0.34 & $0.00632 \pm 0.00064$ & $0.00225 \pm 0.00008$ \\
$\hat{d}_/$        & 0.34 & $0.00269 \pm 0.00044$ & $0.00197 \pm 0.00013$ \\
$\hat{d}_z$        & 0.34 & $0.00341 \pm 0.00162$ & $0.00238 \pm 0.00035$ \\
\end{tabular}
\end{ruledtabular}
\end{table}

\begin{table}
\caption{\label{d_calc}
Calculated $d_\textrm{\small{drag}}$, $\hat{d}_\alpha$, $\hat{d}_/$, and $\hat{d}_z$
for $z = 0.56$ and $z = 0.34$, as a function of $\Omega_m$,
for $\Omega_k = 0$ and $\Omega_{de}(a) \equiv \Omega_\Lambda$ constant.
$d_\textrm{\small{drag}}$ is the BAO galaxy comoving 
standard ruler length in units of $c/H_0$. It is
calculated from $d_\textrm{\small{drag}} = 1.0184 d_*$, $d_* \equiv \theta_* \chi(z_*)$,
$\theta_* = 0.0104092$, $\chi \equiv \int_0^{z_*} dz/E(z)$,
$E(a)=(\Omega_m / a^3 + \Omega_r / a^4 + \Omega_\Lambda + \Omega_k / a^2)^{1/2}$, 
and $a = 1/(1+z)$.
$\hat{d}_\alpha$, $\hat{d}_/$, and $\hat{d}_z$ are 
calculated with equations (5), (6), and (7) of [1] with $z_c = 3.79$.
The dependence on $h = 0.7$ or $\sum m_\nu = 0.27$ eV is negligible
compared to the uncertainties in Table \ref{uncertainties}.
}
\begin{ruledtabular}
\begin{tabular}{c|c|ccc|ccc}
$\Omega_m$ & $100 d_\textrm{\small{drag}}$ & $100 \hat{d}_\alpha$ & $100 \hat{d}_/$ & $100 \hat{d}_z$ 
  & $100 \hat{d}_\alpha$ & $100 \hat{d}_/$ & $100 \hat{d}_z$ \\
\hline
 & & \multicolumn{3}{c|}{$z = 0.56$} & \multicolumn{3}{c}{$z = 0.34$} \\
0.25 & 3.628 & 3.535 & 3.510 & 3.477 & 3.560 & 3.538 & 3.510 \\
0.27 & 3.519 & 3.457 & 3.444 & 3.427 & 3.471 & 3.457 & 3.440 \\
0.28 & 3.468 & 3.421 & 3.414 & 3.405 & 3.429 & 3.420 & 3.408 \\
0.29 & 3.420 & 3.386 & 3.385 & 3.384 & 3.390 & 3.385 & 3.377 \\
0.31 & 3.330 & 3.323 & 3.333 & 3.346 & 3.317 & 3.319 & 3.321 \\
0.33 & 3.248 & 3.265 & 3.285 & 3.311 & 3.251 & 3.259 & 3.271 \\
\end{tabular}
\end{ruledtabular}
\end{table}

\begin{table}
\caption{\label{Dxpec}
Study of peculiar motion corrections to be added to
the G-G measurements of $\hat{d}_\alpha$, $\hat{d}_/$, and $\hat{d}_z$
in Table 1, obtained from simulations.
}
\begin{ruledtabular}
\begin{tabular}{ccccc}
$z$ & Simulation & $\Delta \hat{d}_\alpha$ & $\Delta \hat{d}_/$ & $\Delta \hat{d}_z$ \\
\hline
0.5 & correct $P(k)$              & 0.000062 & 0.000080 & 0.000112 \\
0.5 & correct $P_\textrm{gal}(k)$ & 0.000096 & 0.000125 & 0.000175 \\
0.3 & correct $P(k)$              & 0.000063 & 0.000080 & 0.000111 \\
0.3 & correct $P_\textrm{gal}(k)$ & 0.000084 & 0.000107 & 0.000148 \\
\end{tabular}
\end{ruledtabular}
\end{table}

\begin{table}
\caption{\label{uncertainties}
Uncertainties of $\hat{d}_\alpha$, $\hat{d}_/$, and $\hat{d}_z$
at 68\% confidence.
}
\begin{ruledtabular}
\begin{tabular}{cccc}
                                   & $\hat{d}_\alpha$ & $\hat{d}_/$ & $\hat{d}_z$ \\
\hline
Method                             & $\pm 0.00003$ & $\pm 0.00004$ & $\pm 0.00008$ \\
Peculiar motion correction         & $\pm 0.00004$ & $\pm 0.00004$ & $\pm 0.00005$ \\
Cosmological \textit{et al.} $+$ & & & \\
statistical fluctuations           & $\pm 0.00029$ & $\pm 0.00055$ & $\pm 0.00070$ \\
\hline
Total                              & $\pm 0.00030$ & $\pm 0.00055$ & $\pm 0.00071$ \\
\end{tabular}
\end{ruledtabular}
\end{table}

Uncertainties of $\hat{d}_\alpha$, $\hat{d}_/$, and $\hat{d}_z$ are presented
in Table \ref{uncertainties}. These uncertainties are dominated by cosmological
and statistical fluctuations, and are estimated from 
the root-mean-square fluctuations of many
measurements, from the width of the distribution of $Q$,
and from the issues discussed in the appendices.

Fits to the two independent selected triplets 
$\hat{d}_\alpha$, $\hat{d}_/$, and $\hat{d}_z$
indicated by a ``$*$" in Table \ref{d_meas}, with the uncertainties in Table \ref{uncertainties},
are presented in Table \ref{galaxy_fits}.

Four Scenarios are considered.
In Scenario 1 the dark energy density
is constant, i.e. $\Omega_\textrm{\small{de}}(a) = \Omega_\Lambda$.
In Scenario 2 the observed acceleration of the expansion of the universe is due
to a gas of negative pressure with an equation of state $w \equiv p/\rho < 0$.
We allow the index $w$ to be a function of $a$ \cite{Chevallier, Linder}:
$w(a) = w_0 + w_a (1 - a)$. Scenario 3 is the same as Scenario 2, except that
$w$ is constant, i.e. $w_a = 0$. In Scenario 4 we assume 
$\Omega_\textrm{\small{de}}(a) = \Omega_\textrm{\small{de}} [1 + w_1 (1 - a)]$.

Note in Table \ref{galaxy_fits} that $\Omega_k$ is consistent with
zero, and $\Omega_{\textrm{de}}(a)$ is consistent
with being independent of the expansion parameter $a$.
For $\Omega_k = 0$ and $\Omega_\textrm{\small{de}}(a) \equiv \Omega_\Lambda$
constant we obtain from Table \ref{galaxy_fits}:
\begin{equation}
\Omega_m = 0.288 \pm 0.037,
\label{Om_gal}
\end{equation}
with $\chi^2 = 1.0$ for 4 degrees of freedom.

\begin{table*}
\caption{\label{galaxy_fits}
Cosmological parameters obtained from the 6 independent galaxy BAO measurements 
indicated with a ``$*$" in Table \ref{d_meas}
in several scenarios. Corrections for peculiar motions are given by
Eq. (\ref{correction}) except, for comparison, the fit ``1*" which has no correction.
Scenario 1 has $\Omega_\textrm{\small{de}}(a)$ constant.
Scenario 3 has $w = w_0$.
Scenario 4 has $\Omega_\textrm{\small{de}}(a) = \Omega_\textrm{\small{de}} \left[1 + w_1 (1 - a)\right]$.
}
\begin{ruledtabular}
\begin{tabular}{c|cccccc}
   & Scenario 1* & Scenario 1 & Scenario 1 & Scenario 3 & Scenario 4 & Scenario 4 \\
\hline
$\Omega_k$ & $0$ fixed  & $0$ fixed  & $0.267 \pm 0.362$ & $0$ fixed & $0$ fixed
  & $0.262 \pm 0.383$ \\
$\Omega_\textrm{\small{de}} + 0.6 \Omega_k$ & $0.712 \pm 0.037$ & $0.712 \pm 0.037$ & $0.738 \pm 0.050$
  & $0.800 \pm 0.364$ & $0.760 \pm 0.151$ & $0.745 \pm 0.148$ \\
$w_0$ & n.a. & n.a. & n.a. & $-0.76 \pm 0.65$ & n.a. & n.a. \\
$w_1$ & n.a. & n.a. & n.a. & n.a. & $0.71 \pm 2.00$ & $0.13 \pm 2.77$ \\
$100 d_\textrm{\small{drag}}$ & $3.48 \pm 0.06$ & $3.487 \pm 0.052$ & $3.48 \pm 0.06$
  & $3.43 \pm 0.16$ & $3.42 \pm 0.19$ & $3.48 \pm 0.21$ \\
$\chi^2/$d.f. & $0.9/4$ & $1.0/4$ & $0.4/3$ & $0.9/3$ & $0.9/3$ & $0.4/2$ \\
\end{tabular}
\end{ruledtabular}
\end{table*}

Final calculations are done with fits and numerical integrations.
Never-the-less, it is convenient to present approximate analytical expressions
obtained from the numerical integrations for the case of flat space and
a cosmological constant.
At decoupling, $z_* = 1089.92 \pm 0.25$ from the Planck ``TT,TE,EE$+$lowE$+$lensing"
measurement \cite{Planck}.
The ``angular distance" at decoupling is $D_A(z_*) \equiv \chi(z_*) a_* c / H_0$, with
\begin{equation}
\chi(z_*) = 3.2675 \left( \frac{h+0.35 \sum{m_\nu}}{0.7} \right)^{0.01}
  \left( \frac{0.28}{\Omega_m} \right)^{0.4},
\label{X}
\end{equation}
which has negligible dependence on $h$ or $\sum{m_\nu}$.

From the Planck ``TT,TE,EE$+$lowE$+$lensing"
measurement \cite{Planck}, $\theta_* = 0.0104092 \pm 0.0000031$.
Then the comoving sound horizon at decoupling is $r_* \equiv d_* c / H_0$, with
\begin{equation}
d_* = \theta_* \chi(z_*) = 0.03401 \left( \frac{0.28}{\Omega_m} \right)^{0.4}.
\label{dstar}
\end{equation}

The BAO standard ruler for galaxies $r_\textrm{\small{drag}}$ is larger than
$r_*$ because last scattering of electrons
occurs after last scattering of photons due to their different number
densities. In the present analysis, we take 
$r_\textrm{\small{drag}} \equiv d_\textrm{\small{drag}} c / H_0$ with
\begin{equation}
\frac{d_\textrm{\small{drag}}}{d_*} = 1.0184 \pm 0.0004,
\label{ddrag_dstar}
\end{equation}
from the Planck ``TT,TE,EE$+$lowE$+$lensing" analysis, with
the uncertainty from 
Equation (10) of Reference \cite{Planck}. Note from 
(\ref{dstar}) and Equation (10) of Reference \cite{Planck}
that (\ref{ddrag_dstar}) is insensitive to cosmological parameters,
so the uncalibrated analysis decouples from $h$ or $\sum m_\nu$.

We can test (\ref{ddrag_dstar}) experimentally. From Table \ref{galaxy_fits}
we obtain $d_\textrm{\small{drag}} = 0.03487 \pm 0.00052$. From (\ref{dstar}) 
and (\ref{Om_gal}) we obtain $d_* = 0.03363 \pm 0.00174$, so
the measured $d_\textrm{\small{drag}} / d_* = 1.037 \pm 0.056$.

To the 6 independent galaxy BAO measurements, we add
the sound horizon angle
$\theta_*$, and
obtain the results presented in Table \ref{galaxy_theta_fits}.
Note that measurements are consistent
with flat space and a cosmological constant. 
Note also that the constraint on $\Omega_k$ becomes tighter
if $\Omega_{\textrm{de}}(a)$ is assumed constant, and
that the constraint on $\Omega_{\textrm{de}}(a)$ becomes tighter
if $\Omega_k$ is assumed zero. 
In the scenario of flat space and a cosmological constant we
obtain 
\begin{equation}
\Omega_m = 0.2724 \pm 0.0047,
\label{Om_gal_theta}
\end{equation}
with $\chi^2 = 1.2$ for 5 degrees of freedom.
This is the final result of the present analysis.

Adding two measurements in the quasar Lyman-alpha 
forest \cite{BH_ijaa, lyman, lyman2} we obtain the
results presented in Table \ref{galaxy_Lya_theta_fits}.
In particular, for flat space and a cosmological constant we obtain
\begin{equation}
\Omega_m = 0.2714 \pm 0.0047.
\label{Om_gal_theta_Lya}
\end{equation}
with $\chi^2 = 10.0$ for 7 degrees of freedom.
Note that the Lyman-alpha measurements tighten
the constraints on $\Omega_k$, $w_0$, $w_1$, and $w_a$.

\begin{figure}
\begin{center}
\scalebox{0.465} {\includegraphics{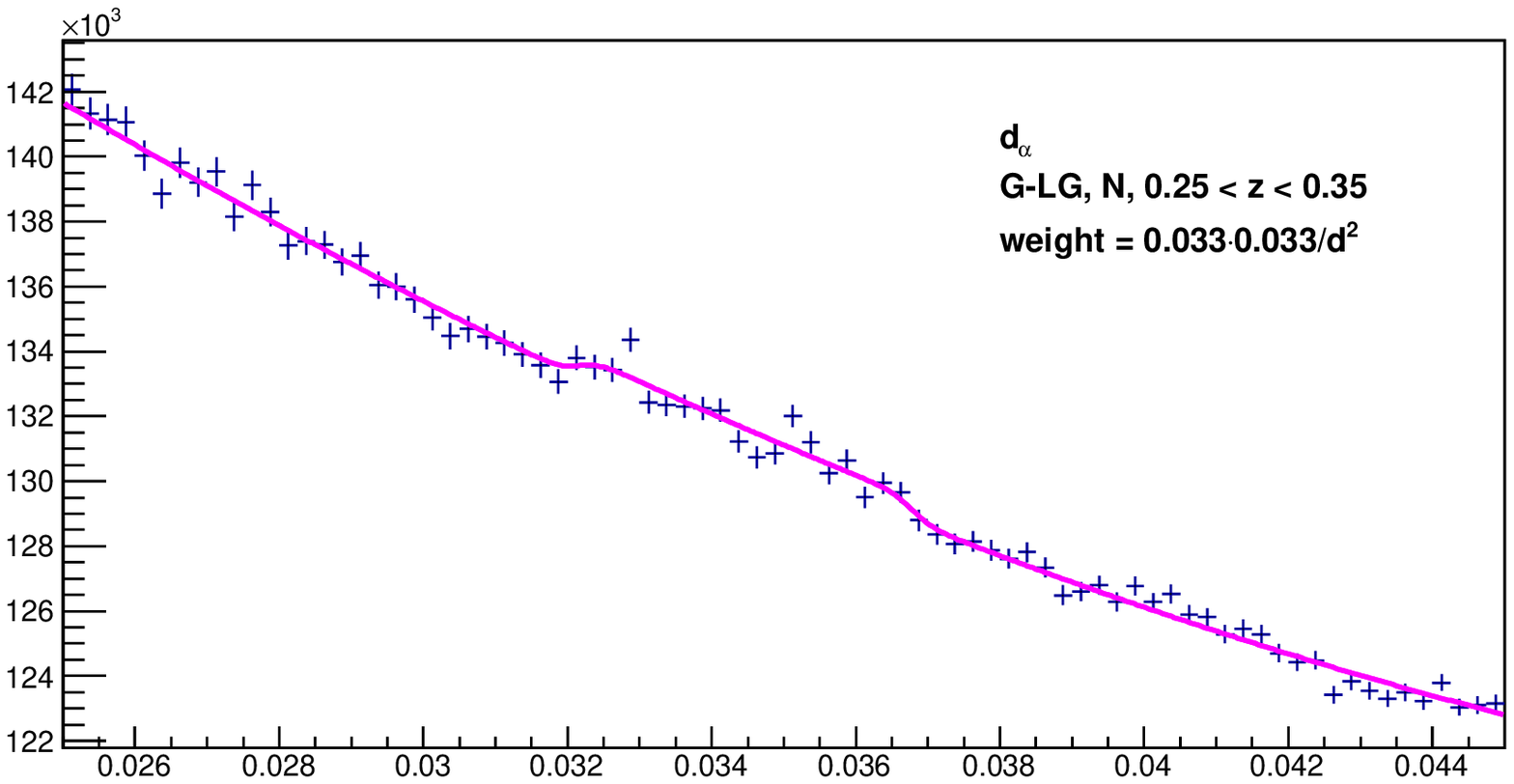}}
\scalebox{0.465} {\includegraphics{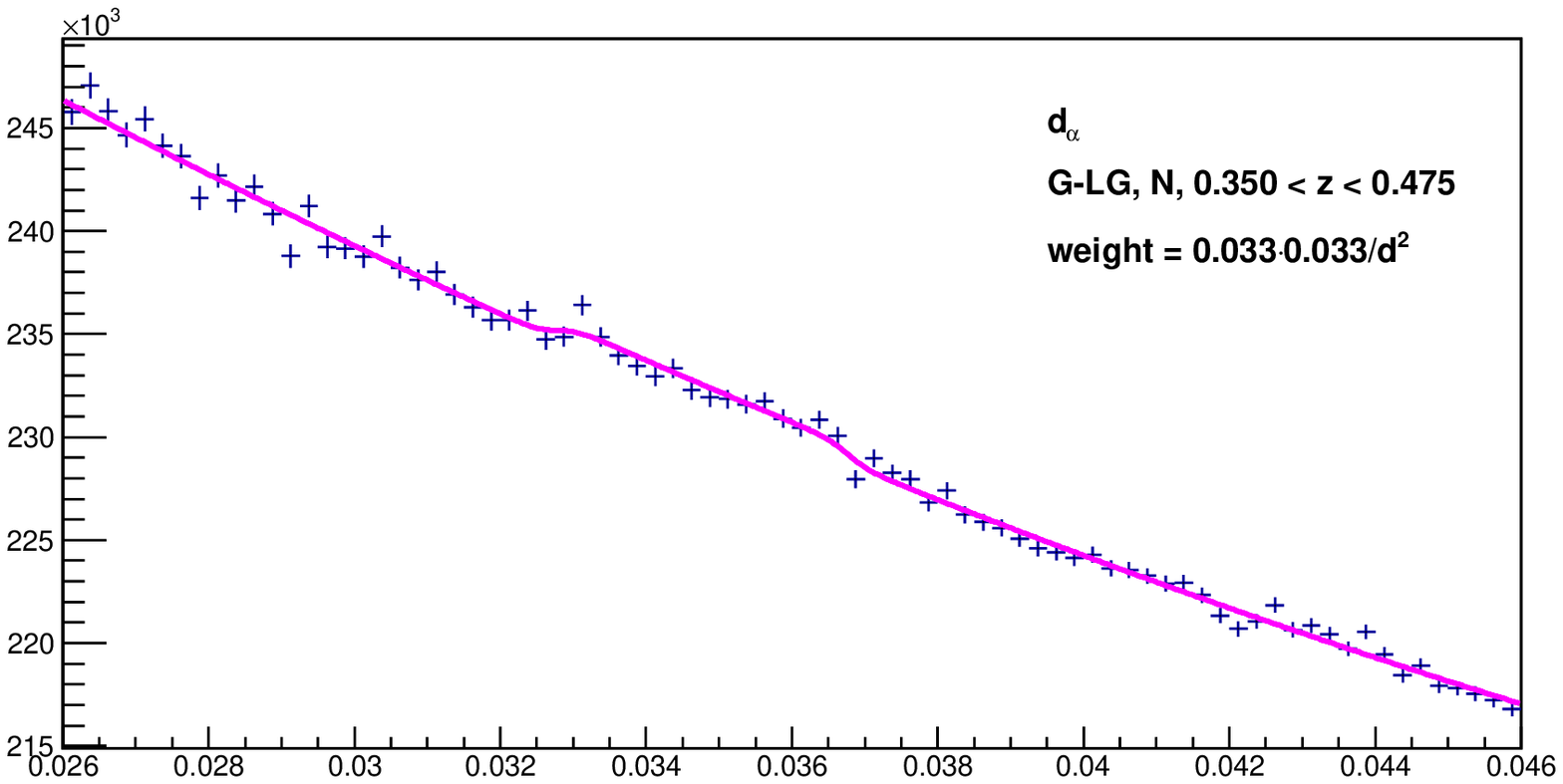}}
\scalebox{0.465} {\includegraphics{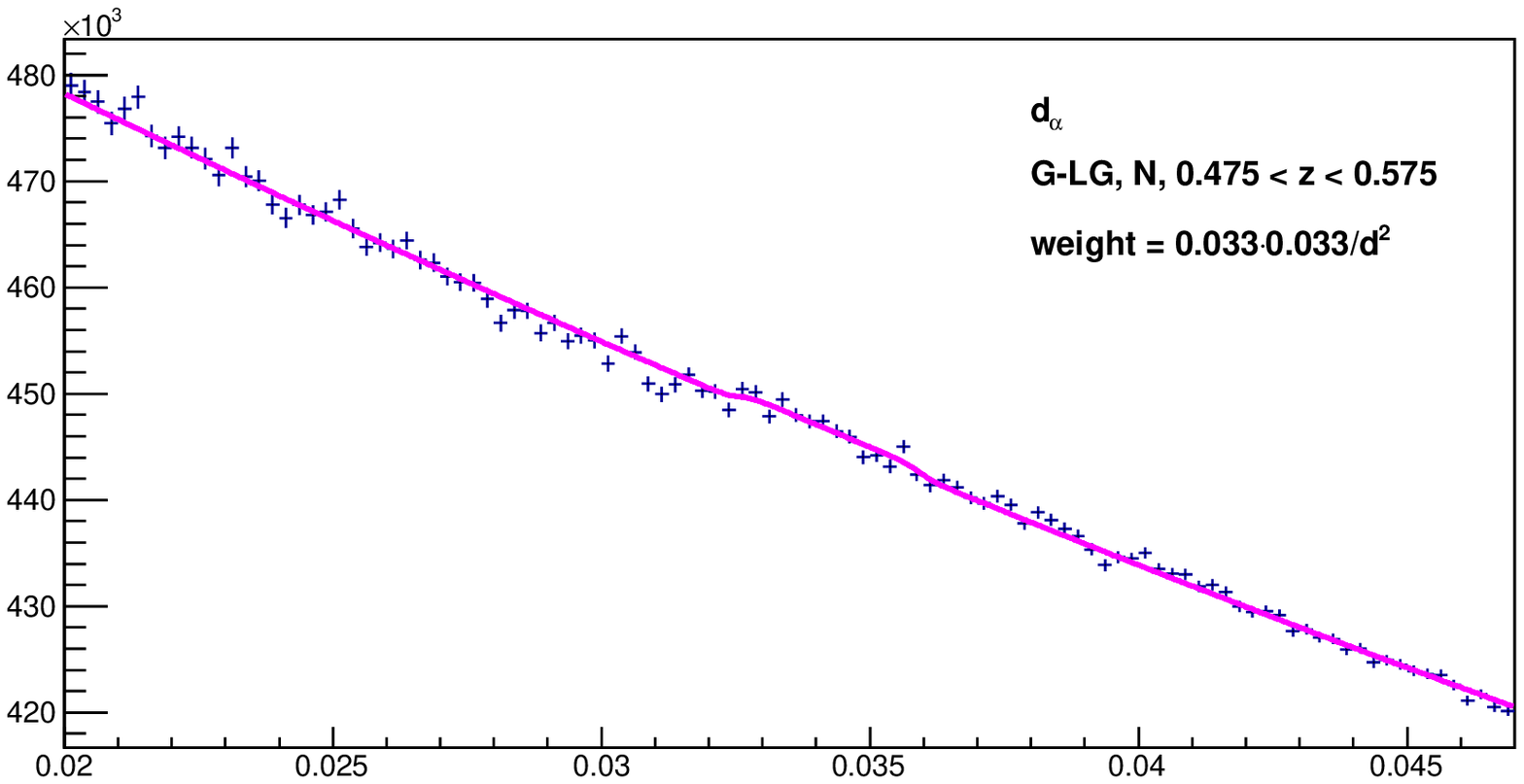}}
\scalebox{0.465} {\includegraphics{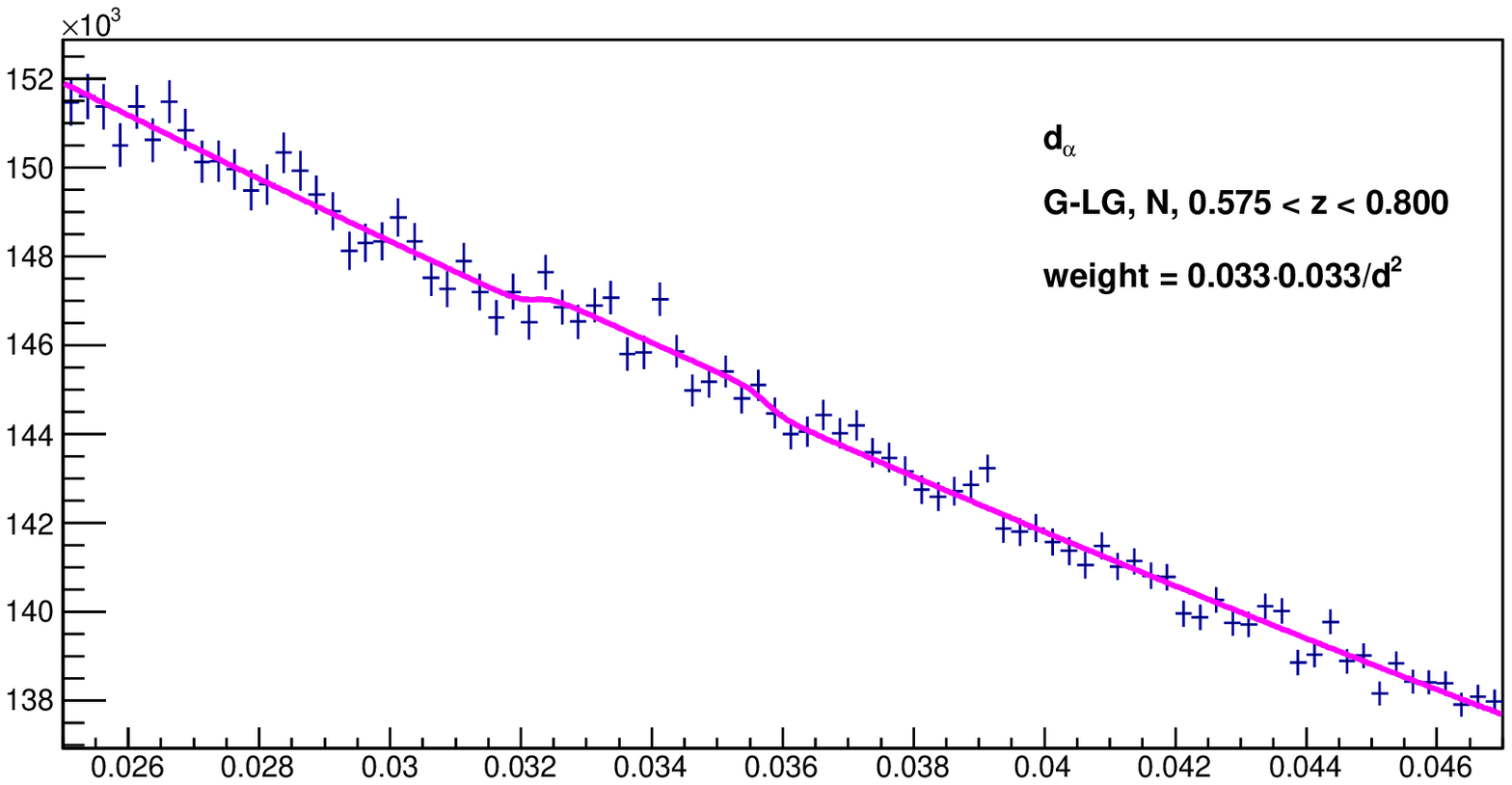}}
\caption{
Fits to histograms of G-LG distances $d$ that obtain
$\hat{d}_\alpha$ at $z = 0.32, 0.42, 0.52$, and $0.65$.
The bins of $z$ are $(0.25, 0.35), (0.35, 0.475), (0.475, 0.575),
\textrm{ and } (0.575, 0.800)$, respectively.
The fits obtain
$\hat{d}_\alpha = 0.03447 \pm 0.00012, 0.03478 \pm 0.00012,
0.03424 \pm 0.00015,
\textrm{ and } 0.03399 \pm 0.00020$
respectively, where uncertainties are statistical from the fits.
A fit with these four measurements (with the total uncertainties of
Table \ref{uncertainties}), plus $\theta_*$ from
the Planck experiment, obtains $\Omega_m = 0.2745 \pm 0.0040$
and $d_* = 0.03433 \pm 0.00020$ with $\chi^2 = 3.0$
for 3 degrees of freedom.
}
\label{fig_s20_250_800}
\end{center}
\end{figure}

As a cross-check of the $z$ dependence, 
from the 4 independent fits to $\hat{d}_\alpha$ 
at different redshifts $z$ presented
in Figure \ref{fig_s20_250_800}, plus $\theta_*$, we obtain
\begin{equation}
\Omega_m = 0.2745 \pm 0.0040,
\label{s20_250_800}
\end{equation}
with $\chi^2 = 3.0$ for	3 degrees of freedom,
for flat space and a cosmological constant.

\begin{figure}
\begin{center}
\scalebox{0.465} {\includegraphics{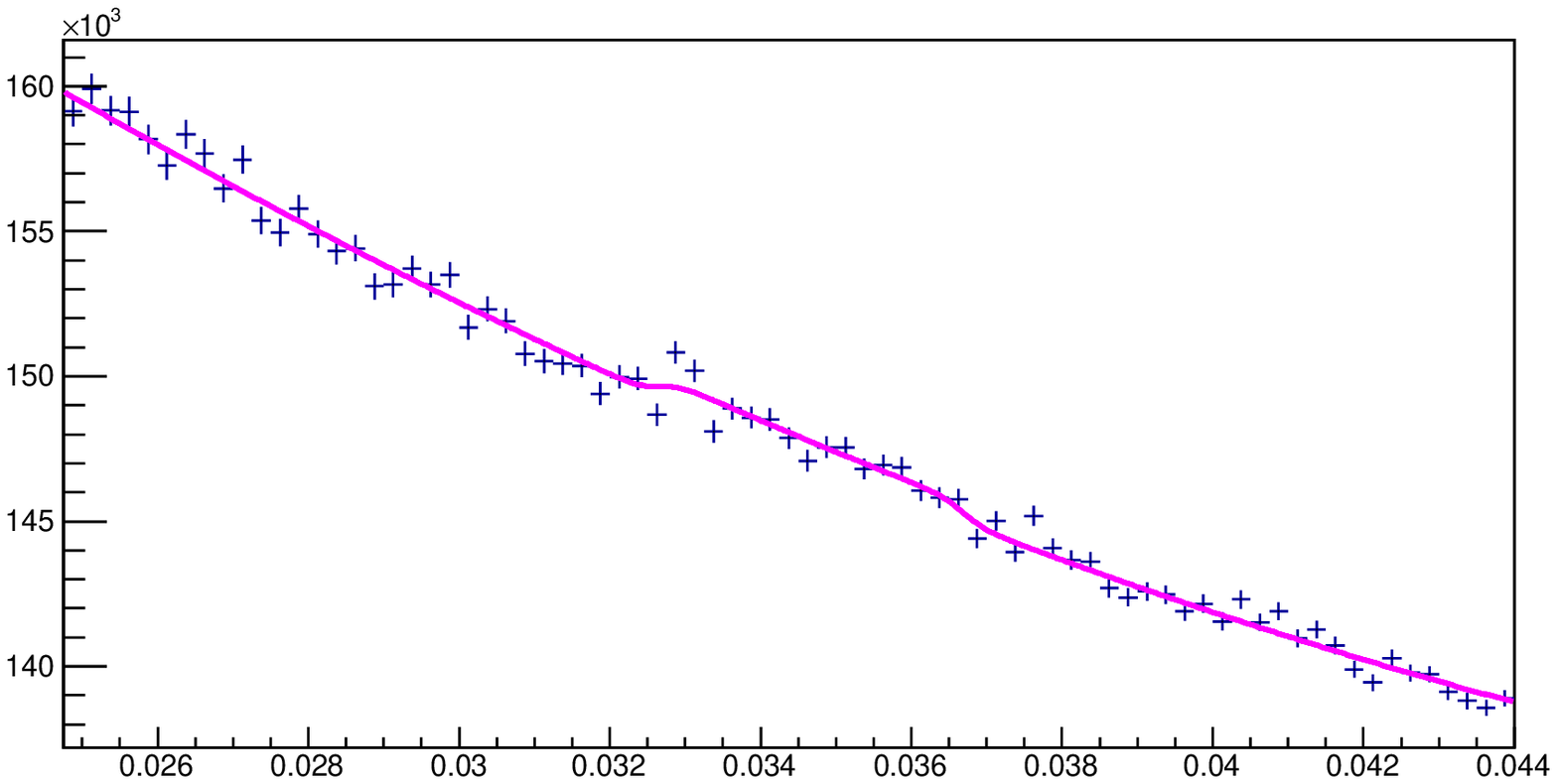}}
\scalebox{0.465} {\includegraphics{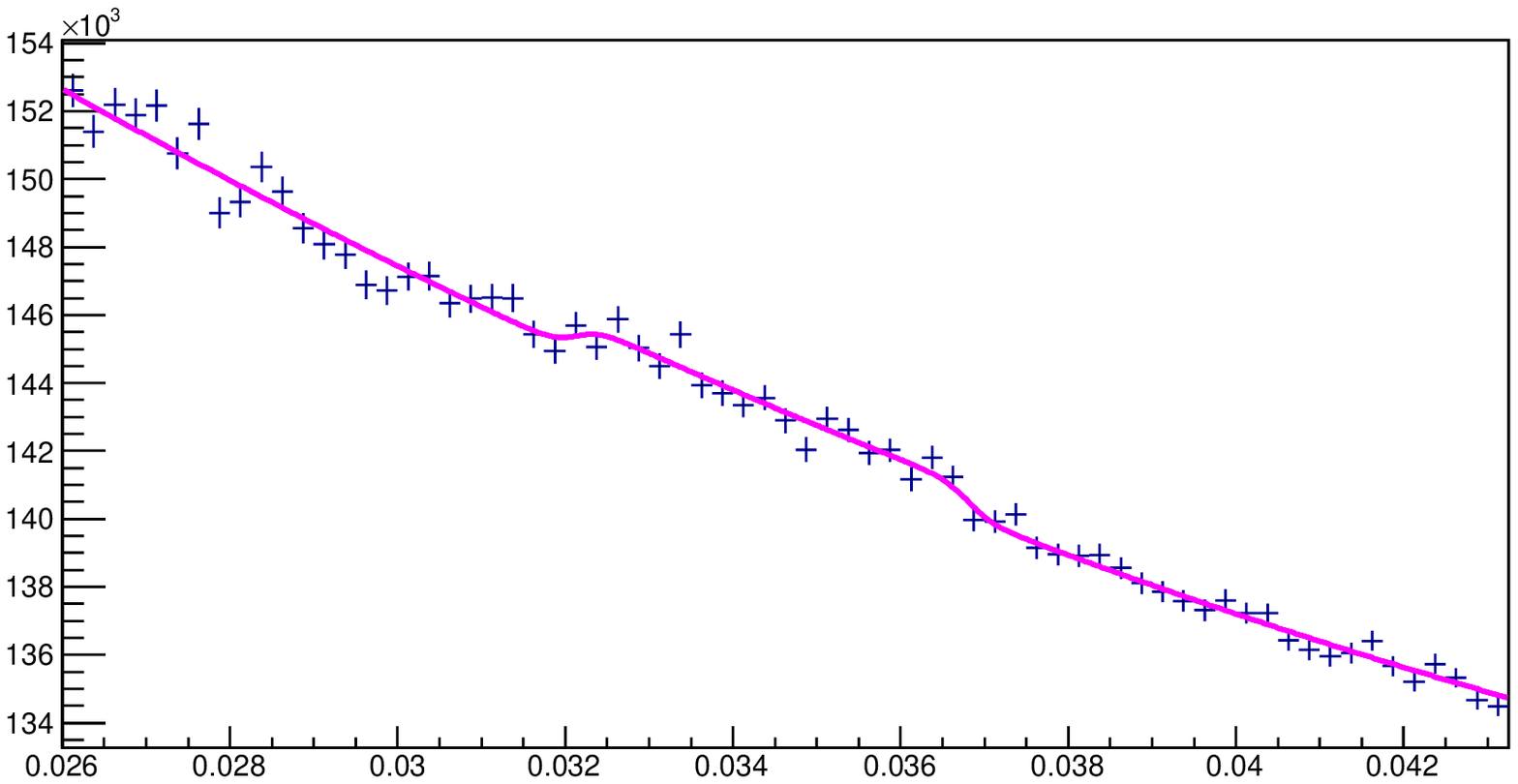}}
\scalebox{0.465} {\includegraphics{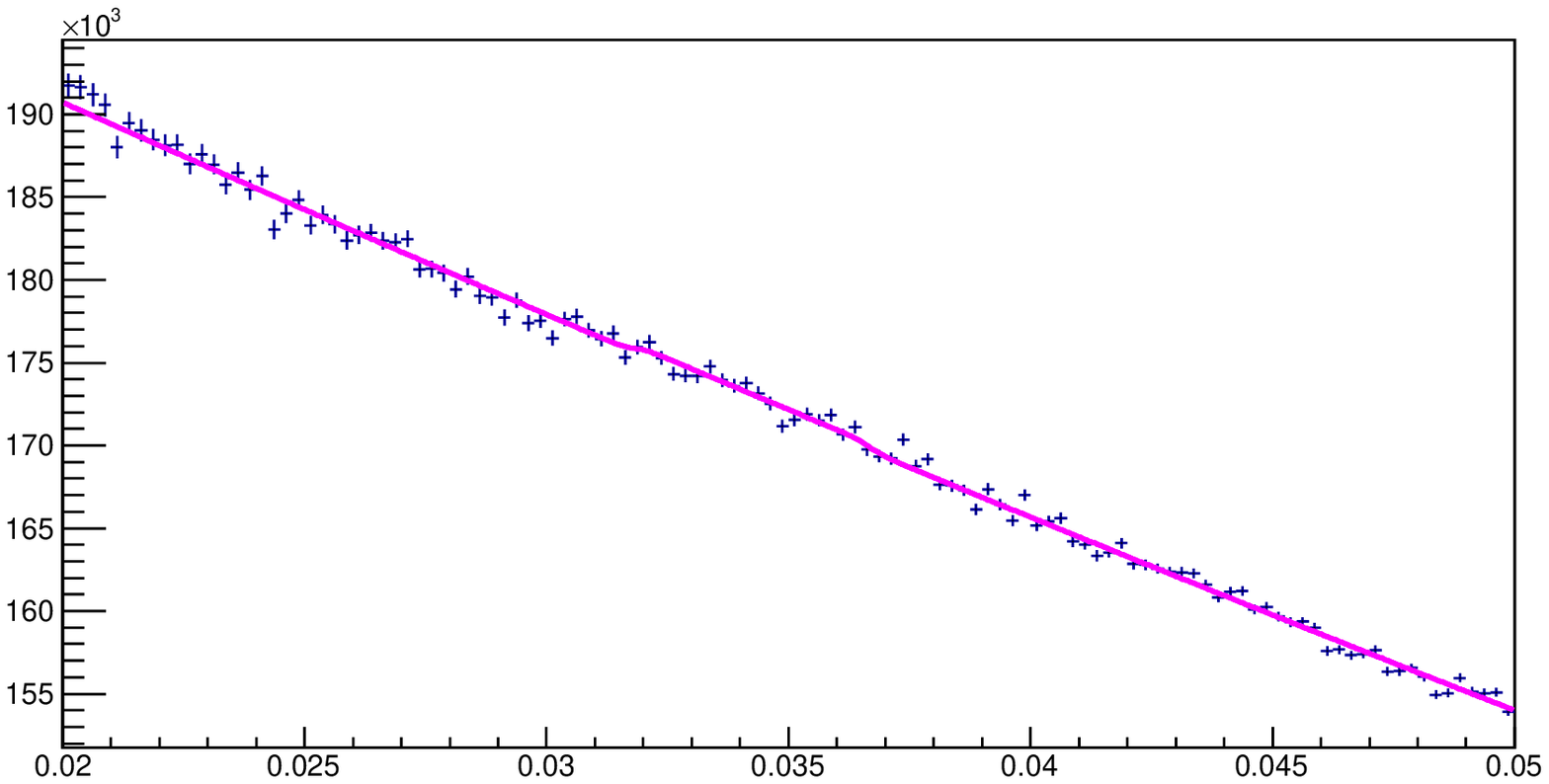}}
\caption{
Fits to histograms of G-LG distances $d$, with $z$ in the range
0.25-0.45, that obtain
$\hat{d}_\alpha$ at $z = 0.36$.
From top to bottom,
they correspond to the northern galactic cap with right ascension $< 180^0$ (NW),
to the northern galactic cap with right ascension $> 180^0$ (NE),
and to the southern galactic cap (S). The fits obtain
$\hat{d}_\alpha = 0.03468 \pm 0.00012, 0.03447 \pm 0.00012, 
\textrm{ and } 0.03424 \pm 0.00019$
respectively, where uncertainties are statistical from the fits. 
A fit with these three measurements (with the total uncertainties of
Table \ref{uncertainties}), plus $\theta_*$ from
the Planck experiment, obtains $\Omega_m = 0.2737 \pm 0.0043$ 
and $d_* = 0.03437 \pm 0.00022$ 
with $\chi^2 = 1.1$ for 2 degrees of freedom. 
}
\label{fig_NE_NW_S}
\end{center}
\end{figure}

As a cross-check of isotropy, from the 3 independent fits to $\hat{d}_\alpha$
at $z = 0.36$ shown in Figure \ref{fig_NE_NW_S}
corresponding to different regions of the sky, we obtain
\begin{equation}
\Omega_m = 0.2737 \pm 0.0043,
\label{NE_NW_S}
\end{equation}
with $\chi^2 = 1.1$ for 2 degrees of freedom,
for flat space and a cosmological constant.

To check the stability of $\hat{d}_\alpha$, $\hat{d}_/$, and $\hat{d}_z$
with the data set and galaxy selections,
we compare fits highlighted with ``$*$" and ``\&"
in Table \ref{d_meas}, and also fits in Figure \ref{fig_250_450_DR13_DR14}.

Additional studies are presented in the appendices.

\begin{table*}
\caption{\label{galaxy_theta_fits}
Cosmological parameters obtained from the 6 independent galaxy BAO measurements 
indicated with a ``$*$" in Table \ref{d_meas},
plus $\theta_*$ from the Planck experiment, in several scenarios.
Corrections for peculiar motions are given by
Eq. (\ref{correction}). $d_\textrm{\small{drag}}/d_* = 1.0184 \pm 0.0004$.
Scenario 1 has $\Omega_\textrm{\small{de}}(a)$ constant.
Scenario 2 has $w(a) = w_0 + w_a (1 - a)$.
Scenario 3 has $w = w_0$.
Scenario 4 has $\Omega_\textrm{\small{de}}(a) = \Omega_\textrm{\small{de}} \left[1 + w_1 (1 - a)\right]$.
}
\begin{ruledtabular}
\begin{tabular}{c|cccccc}
   & Scenario 1 & Scenario 1 & Scenario 2 & Scenario 3 & Scenario 4 & Scenario 4 \\
\hline
$\Omega_k$ & $0$ fixed  & $0.008 \pm 0.018$ & $0$ fixed  &
  $0$ fixed & $0$ fixed & $-0.007 \pm 0.101$ \\
$\Omega_\textrm{\small{de}} + 2.1 \Omega_k$ & $0.7276 \pm 0.0047$ & $0.724 \pm 0.009$ & $0.708 \pm 0.080$
  & $0.724 \pm 0.008$ & $0.723 \pm 0.011$ & $0.723 \pm 0.011$ \\
$w_0$ & n.a. & n.a. & $-0.77 \pm 1.47$ & $-0.95 \pm 0.10$ & n.a. & n.a. \\
$w_a$ or $w_1$ & n.a. & n.a. & $-0.91 \pm 4.53$ & n.a. & $0.19 \pm 0.41$ & $0.35 \pm 2.20$ \\
$100 d_*$ & $3.443 \pm 0.024$ & $3.42 \pm 0.06$ & $3.35 \pm 0.04$
  & $3.41 \pm 0.07$ & $3.41 \pm 0.09$ & $3.39 \pm 0.20$ \\
$\chi^2/$d.f. & $1.2/5$ & $1.0/4$ & $0.9/3$ & $1.0/4$ & $1.0/4$ & $1.0/3$ \\
\end{tabular}
\end{ruledtabular}
\end{table*}

\begin{table*}
\caption{\label{galaxy_Lya_theta_fits}
Cosmological parameters obtained from the 6 galaxy BAO measurements 
indicated with a ``$*$" in Table \ref{d_meas},
plus $\theta_*$ from the Planck experiment, plus 
two Lyman-alpha measurements \cite{BH_ijaa, lyman, lyman2} in several scenarios.
Corrections for peculiar motions are given by
Eq. (\ref{correction}).
$d_\textrm{\small{drag}}/d_* = 1.0184 \pm 0.0004$.
Scenario 1 has $\Omega_\textrm{\small{de}}(a)$ constant.
Scenario 2 has $w(a) = w_0 + w_a (1 - a)$.
Scenario 3 has $w = w_0$.
Scenario 4 has $\Omega_\textrm{\small{de}}(a) = \Omega_\textrm{\small{de}} \left[1 + w_1 (1 - a)\right]$.
}
\begin{ruledtabular}
\begin{tabular}{c|cccccc}
   & Scenario 1 & Scenario 1 & Scenario 2 & Scenario 3 & Scenario 4 & Scenario 4 \\
\hline
$\Omega_k$ & $0$ fixed  & $-0.011 \pm 0.008$ & $0$ fixed  &
  $0$ fixed & $0$ fixed & $-0.022 \pm 0.010$ \\
$\Omega_\textrm{\small{de}} + 2.1 \Omega_k$ & $0.7286 \pm 0.0047$ & $0.734 \pm 0.006$ & $0.703 \pm 0.028$
  & $0.726 \pm 0.008$ & $0.723 \pm 0.011$ & $0.720 \pm 0.011$ \\
$w_0$ & n.a. & n.a. & $-0.70 \pm 0.33$ & $-0.96 \pm 0.09$ & n.a. & n.a. \\
$w_a$ or $w_1$ & n.a. & n.a. & $-1.18 \pm 1.37$ & n.a. & $0.24 \pm 0.40$ & $0.80 \pm 0.49$ \\
$100 d_*$ & $3.449 \pm 0.024$ & $3.48 \pm 0.04$ & $3.32 \pm 0.13$
  & $3.42 \pm 0.07$ & $3.40 \pm 0.08$ & $3.34 \pm 0.09$ \\
$\chi^2/$d.f. & $10.0/7$ & $7.7/6$ & $8.0/5$ & $9.2/6$ & $9.0/6$ & $4.6/5$ \\
\end{tabular}
\end{ruledtabular}
\end{table*}

\section{Measurement of $H_0$ with BAO as a \textit{calibrated} standard ruler}


We consider the scenario of flat space
and a cosmological constant.
It is useful to present approximate analytic expressions,
tho all final calculations are done directly with fits 
to the measurements marked with a ``$*$" in Table \ref{d_meas} and 
numerical integrations to obtain correct uncertainties for
correlated parameters.
To calibrate the BAO measurements, we integrate the
comoving photon-electron-baryon plasma sound speed from $t = 0$ up to
decoupling and obtain the ``comoving acoustic horizon distance" 
$r_* \equiv d_* c/H_0$, with
\begin{eqnarray}
d_* & = & 0.03407 \left( \frac{h+0.026 \sum{m_\nu}}{0.7} \right)^{0.513} \nonumber \\
 & & \times \left( \frac{0.28}{\Omega_m} \right)^{0.244} \left( \frac{0.0225}{\Omega_b h^2} \right)^{0.097}. 
\label{d_star}
\end{eqnarray}
The acoustic angular scale is
\begin{eqnarray}
\theta_* \equiv \frac{d_*}{\chi(z_*)}  
& = & 0.010427 \left( \frac{h+0.020 \sum m_{\nu}}{0.70} \right)^{0.503} \nonumber \\
 & &  \times \left( \frac{\Omega_m}{0.28} \right)^{0.156} \left( \frac{0.0225}{\Omega_b h^2} \right)^{0.097},
\label{theta}
\end{eqnarray}
in agreement with Equation (11) of \cite{Planck}.

Let us now consider the measurement of $h$.
From the galaxy BAO measurements in Table \ref{galaxy_fits}
we obtain $\Omega_m = 0.288 \pm 0.037$ and $d_\textrm{\small{drag}} = 0.03487 \pm 0.00052$.
From Big Bang Nucleosynthesis, $\Omega_b h^2 = 0.0225 \pm 0.0008$ 
at 68\% confidence \cite{PDG2018}.
From this data and Equations (\ref{ddrag_dstar}) and (\ref{d_star}),
or the corresponding fit, we obtain
\begin{equation}
h + 0.026 \sum{m_\nu} = 0.716 \pm 0.027,
\label{h_galaxies}
\end{equation}
with $\chi^2 = 1.0$ for 4 degrees of freedom.

The Planck measurement of $\theta_*$ allows a more precise measurement of $h$.
From Table \ref{galaxy_theta_fits} we obtain 
$\Omega_m = 0.2724 \pm 0.0047$. Then from Big Bang Nucleosynthesis and
(\ref{theta}), or the corresponding fit, we obtain
\begin{equation}
h + 0.020 \sum{m_\nu} = 0.7038 \pm 0.0060,
\label{h_galaxies_theta}
\end{equation}
with $\chi^2 = 1.2$ for	5 degrees of freedom.
Note that the uncertainties of $h$ and $\Omega_m$ are correlated through
Equation (\ref{theta}).

\section{Studies of CMB fluctuations}

\begin{figure}
\begin{center}
\scalebox{0.465} {\includegraphics{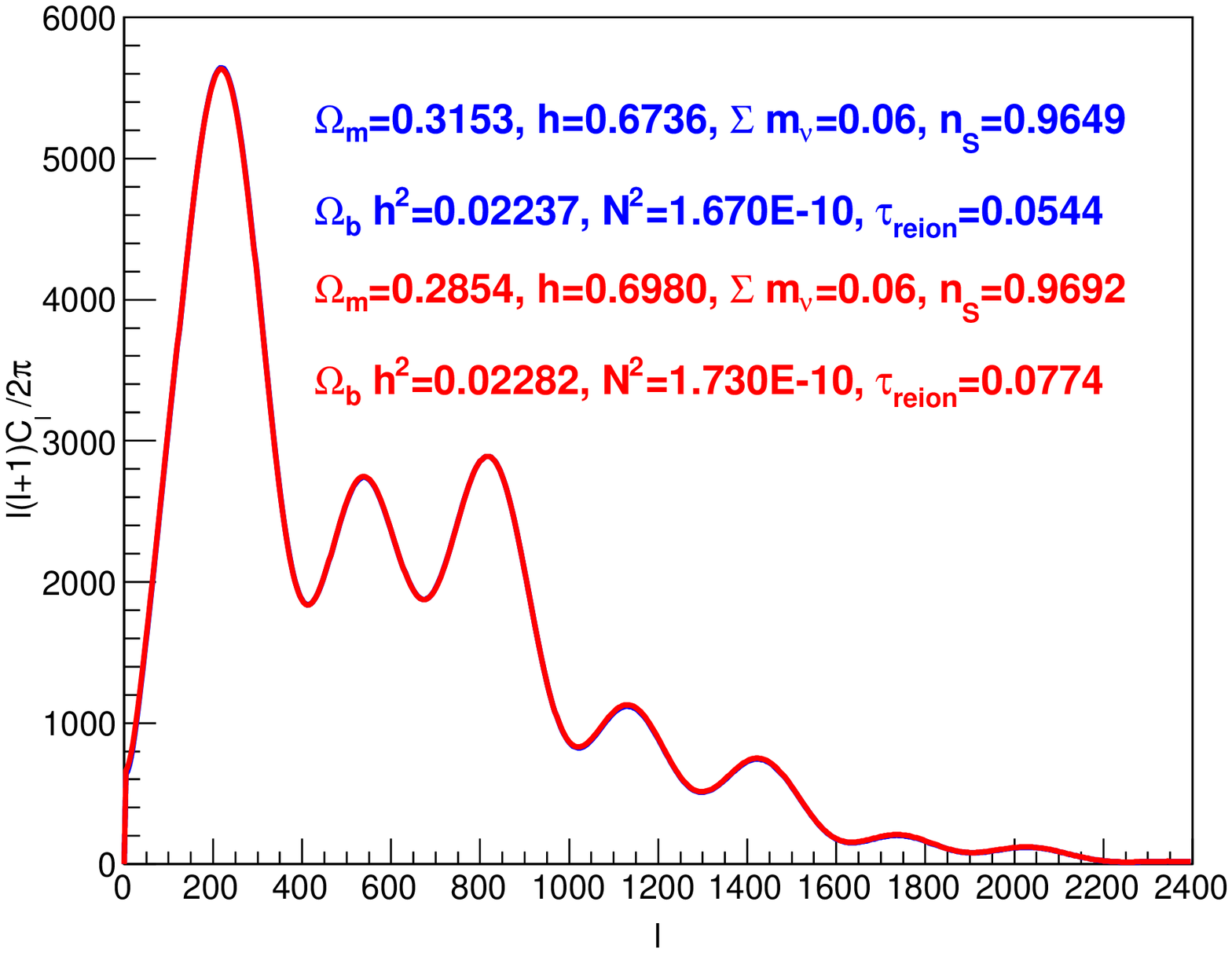}}
\caption{
Comparison of the power spectra $l (l+1) C_{TT,l}^S / ( 2 \pi )$
$[\mu \textrm{K}^2]$ for the
Planck 2018 ``TT,TE,EE$+$lowE$+$lensing" parameters, with the
best fit spectra with $\Omega_m = 0.2854$ and $\sum m_\nu = 0.06$ eV
fixed, calculated with the approximate Equation (7.2.41) of \cite{Weinberg}
(modified to include $\sum m_\nu$).
The r.m.s. difference is $6.07 \mu \textrm{K}^2$, corresponding to 0.11\% of the
first acoustic peak, so the two spectra can not be distinguished by eye.
}
\label{Cl}
\end{center}
\end{figure}

In Table \ref{comparisons} we present a qualitative study
of the sensitivity of the CMB power spectrum
$l (l+1) C_{TT,l}^S / ( 2 \pi )$ to constrain $\Omega_m$ and $\sum m_\nu$.
We use the approximate analytic expression (7.2.41) of \cite{Weinberg},
modified to include $\sum m_\nu$,
to compare the spectra with Planck 2018 ``TT,TE,EE$+$lowE$+$lensing" 
parameters with the best fit spectra with fixed values 
$\Omega_m = 0.2854$ and $\sum m_\nu = 0.06, 0.1, 0.2, 0.3, 0.4, 0.5$ eV.
We find that the differences in spectra range from 0.11\% to 0.3\% of
the first acoustic peak, see Figure \ref{Cl}.
So the CMB power spectrum, while being very sensitive to constrain $\theta_*$,
has low sensitivity to constrain $\Omega_m$ or $\sum m_\nu$.

\begin{table}
\centering
\caption{\label{comparisons}
Cosmologies with fixed $\Omega_m$
and $\sum m_\nu$ fitted to the
CMB power spectrum 
$l (l+1) C_{TT,l}^S / ( 2 \pi )$ with the
Planck 2018 ``TT,TE,EE$+$lowE$+$lensing" parameters
$\Omega_m = 0.3153$,
$\sum m_\nu = 0.06$ eV, $h = 0.6736$, $\Omega_b h^2 = 0.02237$,
$n_s=0.9649$, $N^2 = 1.670 \times10^{-10}$, and
$\tau = 0.0544$ \cite{Planck}. 
The approximate analytic equation (7.2.41) of \cite{Weinberg} 
(modified to include $\sum m_\nu$) was used.
Notation: $N^2 \equiv A_s / (4 \pi ) \equiv \Delta_R^2 / (4 \pi )$.
}
\begin{ruledtabular}
{\begin{tabular}{lcccccc}
$\Omega_m$      & 0.2854 & 0.2854 & 0.2854 & 0.2854 & 0.2854 & 0.2854 \\
$\sum m_\nu$ [eV] & 0.06  & 0.1   & 0.2    & 0.3    & 0.4    & 0.5 \\
\hline
$h$             & 0.6980 & 0.6976 & 0.6965 & 0.6954 & 0.6942 & 0.6931 \\
$100 \Omega_b h^2$  & 2.282 & 2.288 & 2.306 & 2.324 & 2.343 & 2.362 \\
$n_s$           & 0.9692 & 0.9699 & 0.9716 & 0.9735 & 0.9754 & 0.9774  \\
$10^{10} N^2$   & 1.730 & 1.729   & 1.725 & 1.722   & 1.716 & 1.713 \\
$\tau$          & 0.0774 & 0.0778 & 0.0787 & 0.0797 & 0.0799 & 0.0809  \\
\hline
r.m.s. $[\mu \textrm{K}^2]$ & 6.07 & 6.98 & 9.29 & 11.66 & 14.06 & 16.49 \\
\end{tabular}}
\end{ruledtabular}
\label{BAO_fit2}
\end{table}

In view of the low sensitivity of the CMB power spectra to constrain $\Omega_m$,
the Planck analysis can benefit from a combination with the direct
measurement of $\Omega_m$ given by Equation (\ref{Om_gal_theta}). 
The combination, obtained with the
``base\_mnu\_plikHM\_TTTEEE\_lowTEB\_lensing\_*.txt MC chains" 
made public by the Planck Collaboration \cite{Planck},
is presented in Table \ref{combination}. This combination is
preliminary due to the sparseness of the MC chains at low values of
$\Omega_m$.

\begin{table}
\centering
\caption{
Combination of the Planck 2018 ``TT,TE,EE$+$lowE$+$lensing" analysis \cite{Planck} with
the directly measured
$\Omega_m = 0.2724 \pm 0.0047$. Uncertainties are at $68\%$ confidence.
The Planck $\chi^2_\textrm{\tiny{P}} \equiv -2 \cdot \ln \mathcal{L}$ 
increases from 12956.78 to 12968.64 with this combination.
The galaxy $\chi^2_\textrm{\tiny{G}} \equiv (\Omega_m - 0.2724)^2/0.0047^2$.
Preliminary.
}
\begin{ruledtabular}
{\begin{tabular}{l|cc}
                  & Planck                & Planck$+ \Omega_m$ \\
\hline
$\Omega_b h^2$    & $0.02237 \pm 0.00015$ & $0.02265 \pm 0.00012$ \\
$\Omega_c h^2$    & $0.1200 \pm	0.0012$   & $0.1155 \pm 0.0005$ \\
$100 \theta_*$    & $1.04092 \pm 0.00031$ & $1.04125 \pm 0.00022$ \\
$\tau$            & $0.0544 \pm 0.0073$   & $0.078 \pm 0.006$ \\
$\ln 10^{10} A_s$ & $3.044 \pm 0.014$     & $3.102 \pm 0.020$ \\
$n_s$             & $0.9649 \pm 0.0042$   & $0.9726 \pm 0.0017$ \\
\hline
$\Omega_\Lambda$  & $0.6847 \pm 0.0073$   & $0.7147 \pm 0.0040$ \\
$\Omega_m$        & $0.3153 \pm 0.0073$   & $0.2853 \pm 0.0040$ \\
$h$               & $0.6736 \pm 0.0054$   & $0.6990 \pm 0.0030$ \\
$\sigma_8$        & $0.8111 \pm 0.0060$   & $0.8346 \pm 0.0054$ \\ 
\hline
$\chi^2_\textrm{\tiny{P}}$ & 12956.78 & 12968.64 \\
$\chi^2_\textrm{\tiny{G}}$ & 83.31 & 7.53 \\
\hline
$\chi^2_\textrm{tot}$ & 13040.09 & 12976.17 \\
\end{tabular}}
\end{ruledtabular}
\label{combination}
\end{table}

\section{Tensions}
\label{section_tensions}

We consider four direct measurements:
(i) $h = 0.7348 \pm 0.0166$ by the Sh$_0$es Team \cite{Shoes},
(ii) $\sigma_8 \approx \left[ 0.746 \pm 0.012 \textrm{ (stat)} \pm 0.022 \textrm{ (syst)} \right]
(0.3/\Omega_m)^{0.47}$ from the abundance of rich
galaxy clusters \cite{PDG2018, sigma8_clusters},
(iii) $\sigma_8 \approx \left[ 0.745 \pm 0.039 \right] (0.3/\Omega_m)^{0.5}$
from weak gravitational lensing \cite{PDG2018, sigma8_lensing}, and
(iv) $\Omega_m = 0.2724 \pm 0.0047$ from galaxy BAO and $\theta_*$ from Planck,
Equation (\ref{Om_gal_theta}) of this analysis.
Comparing these measurements with Planck (left hand column
of Table \ref{combination}) we obtain differences of $3.5 \sigma$, $2.5 \sigma$, $1.8 \sigma$,
and $4.9 \sigma$, respectively.
Comparing these measurements with the Planck$+\Omega_m$ combination (right hand column
of Table \ref{combination}) we obtain differences of $2.1 \sigma$, $2.3 \sigma$, $1.5 \sigma$,
and $2.1 \sigma$, respectively.
In conclusion, the Planck$+ \Omega_m$ combination
reduces	the tensions with the direct measurements.
Note that the Planck$+\Omega_m$ combination has $\sigma_8$
greater than the direct measurements. This
$2.7 \sigma$ tension may be due to neutrino masses.

\section{Update on neutrino masses}

We consider the scenario of three neutrino flavors with
eigenstates of nearly the same mass, so $\sum m_\nu \approx 3 m_\nu$.

Massive neutrinos suppress the power spectrum of linear density
fluctuations $P(k)$ by a factor $1 - 8 \Omega_\nu / \Omega_m$
for $k >> 0.018 \cdot \Omega_m^{1/2} (\sum m_\nu / 1\textrm{ eV})^{1/2} h$ Mpc$^{-1}$
\cite{f_mnu}. This suppression affects $\sigma_8$ and the galaxy power spectrum 
$P_\textrm{gal}(k)$, 
\textit{but does not} affect the Sachs-Wolfe effect at low $k$.
So, by comparing fluctuations at large and small $k$ it is possible
to constrain or measure $\sum m_\nu$ \cite{BH_ijaa_3}.

To obtain $\sum	m_\nu$ we minimize a $\chi^2$ with
four terms corresponding to $N^2$, $\sigma_8$,
and two parameters obtained from the Planck$+ \Omega_m$
combination:
$h = 0.6990 \pm 0.0030$, and $n_s = 0.9726 \pm 0.0017$.
In the fit, $\Omega_m$ is obtained from Equation (\ref{theta}), 
and $\Omega_b h^2 = 0.02265 \pm 0.00012$.
$\sigma_8$ is obtained from the	combination of the
two direct measurements	presented in Section \ref{section_tensions}.

For $N^2 = (2.08 \pm 0.33) \times 10^{-10}$ \cite{BH_ijaa_3}
obtained from the Sachs-Wolfe effect measured by the COBE satellite 
(see list of references in \cite{Weinberg}) we obtain
\begin{equation}
\sum m_\nu = 0.45 \pm 0.20\textrm{ eV},
\end{equation}
with zero degrees of freedom, in agreement with \cite{BH_ijaa_3}
where the method is explained in detail.

Since $\sum m_\nu < 1.7$ eV, neutrinos are still
ultra-relativistic at decoupling.
Then there is no power suppression of the CMB fluctuations,
and we can use the entire spectrum to fix the amplitude $N^2$.
From the Planck$+ \Omega_m$ combination of Table \ref{combination} we 
obtain $N^2 \equiv A_s / (4 \pi ) = (1.7700 \pm 0.0354) \times 10^{-10}$,
and
\begin{equation}
\sum m_\nu = 0.26 \pm 0.08 \textrm{ eV},
\end{equation}
with zero degrees of freedom.

To strengthen the constraints from the two direct measurements of
$\sigma_8$, we add to the fit measurements of fluctuations
of number counts of galaxies in spheres of radii $16/h, 32/h, 64/h,$ and $128/h$
Mpc, as explained in \cite{BH_ijaa_3}. We obtain
\begin{equation}
\sum m_\nu = 0.27 \pm 0.08 \textrm{ eV},
\label{summnu}
\end{equation}
with $\chi^2 = 1.6$ for 2 degrees of freedom, and
find no significant pulls on $N^2$, $h$, or $n_s$.
These results are sensitive to the accuracy of 
the direct measurements of $\sigma_8$.



\section{Acknowledgment}
We have used data in the publicly released
Sloan Digital Sky Survey SDSS DR14 catalog.

Funding for the Sloan Digital Sky Survey (SDSS) has been provided
by the Alfred P. Sloan Foundation, the Participating Institutions,
the National Aeronautics and Space Administration, the
National Science Foundation, the U.S. Department of Energy, the
Japanese Monbukagakusho, and the Max Planck Society.
The SDSS Web site is http://www.sdss.org/.

The SDSS is managed by the Astrophysical Research Consortium (ARC)
for the Participating Institutions. The Participating Institutions
are The University of Chicago, Fermilab, the Institute for Advanced Study,
the Japan Participation Group, The Johns Hopkins University,
Los Alamos National Laboratory, the Max-Planck-Institute for Astronomy (MPIA),
the Max-Planck-Institute for Astrophysics (MPA), New Mexico State University,
University of Pittsburgh, Princeton University,
the United States Naval Observatory, and the University of Washington.

We have also used data publicly released by the Planck Collaboration
\cite{Planck} in the form of ``MC chains", and the corresponding
analysis tool ``GetDist GUI".


\appendix

\begin{figure}
\begin{center}
\scalebox{0.465} {\includegraphics{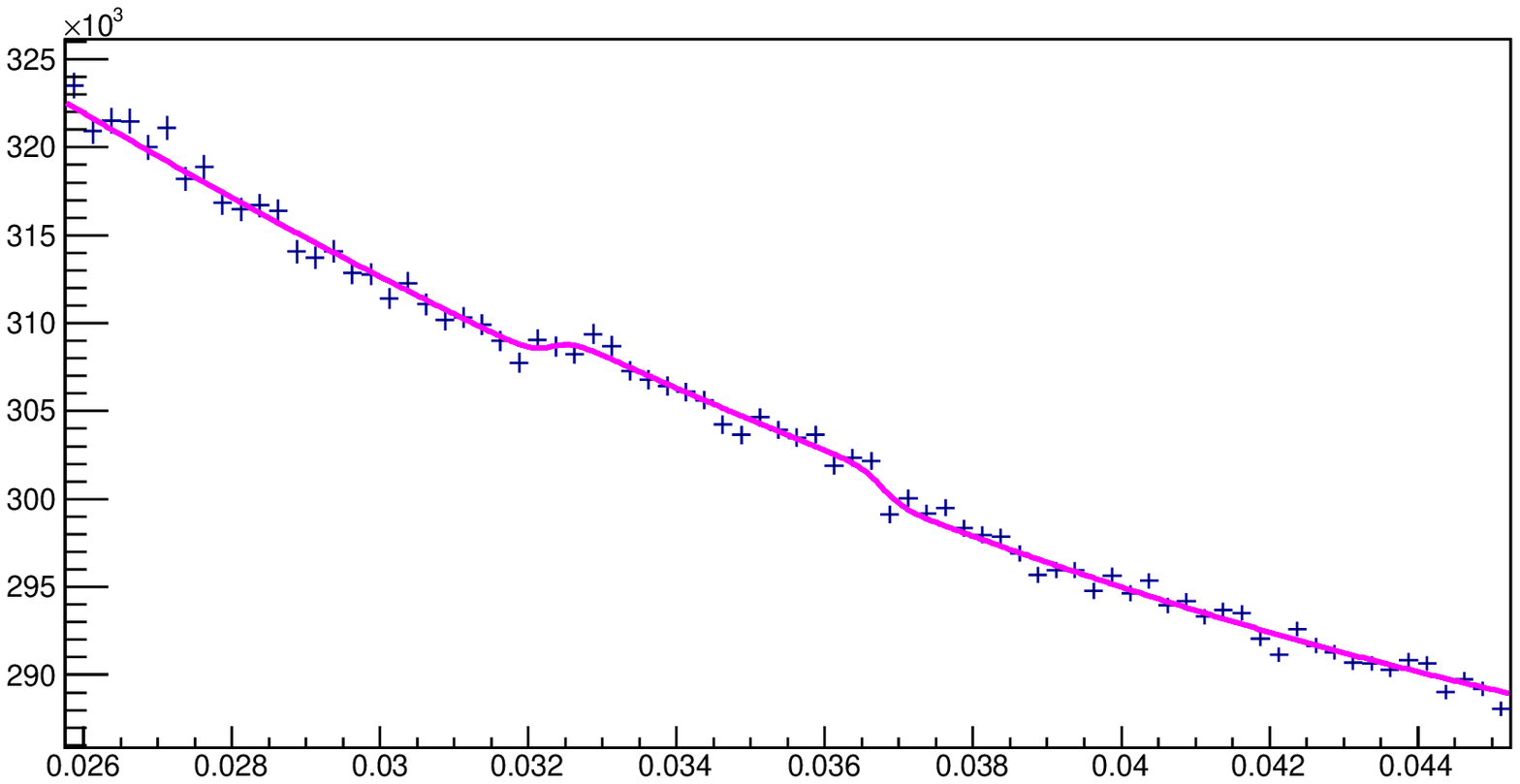}}
\scalebox{0.465} {\includegraphics{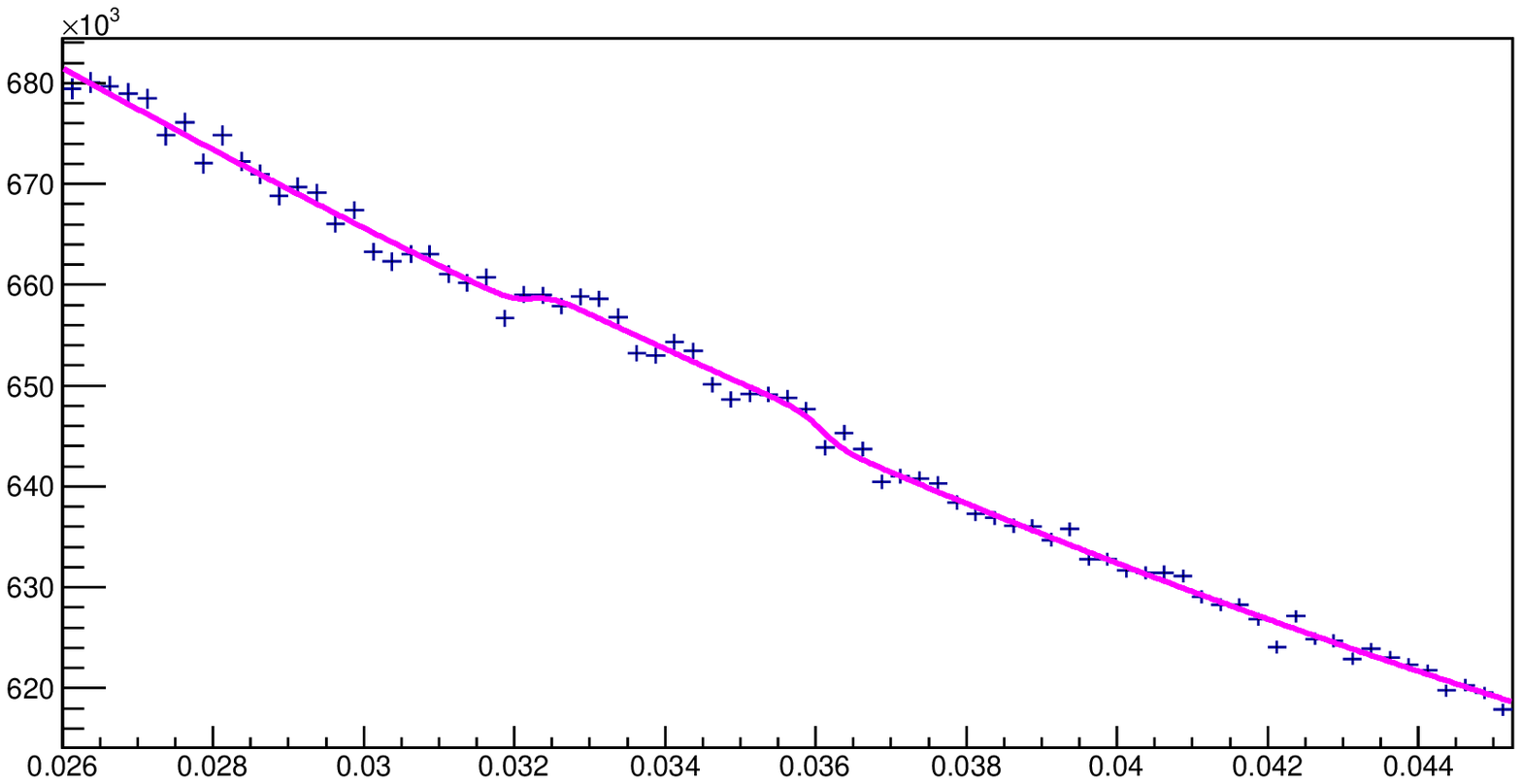}}
\scalebox{0.465} {\includegraphics{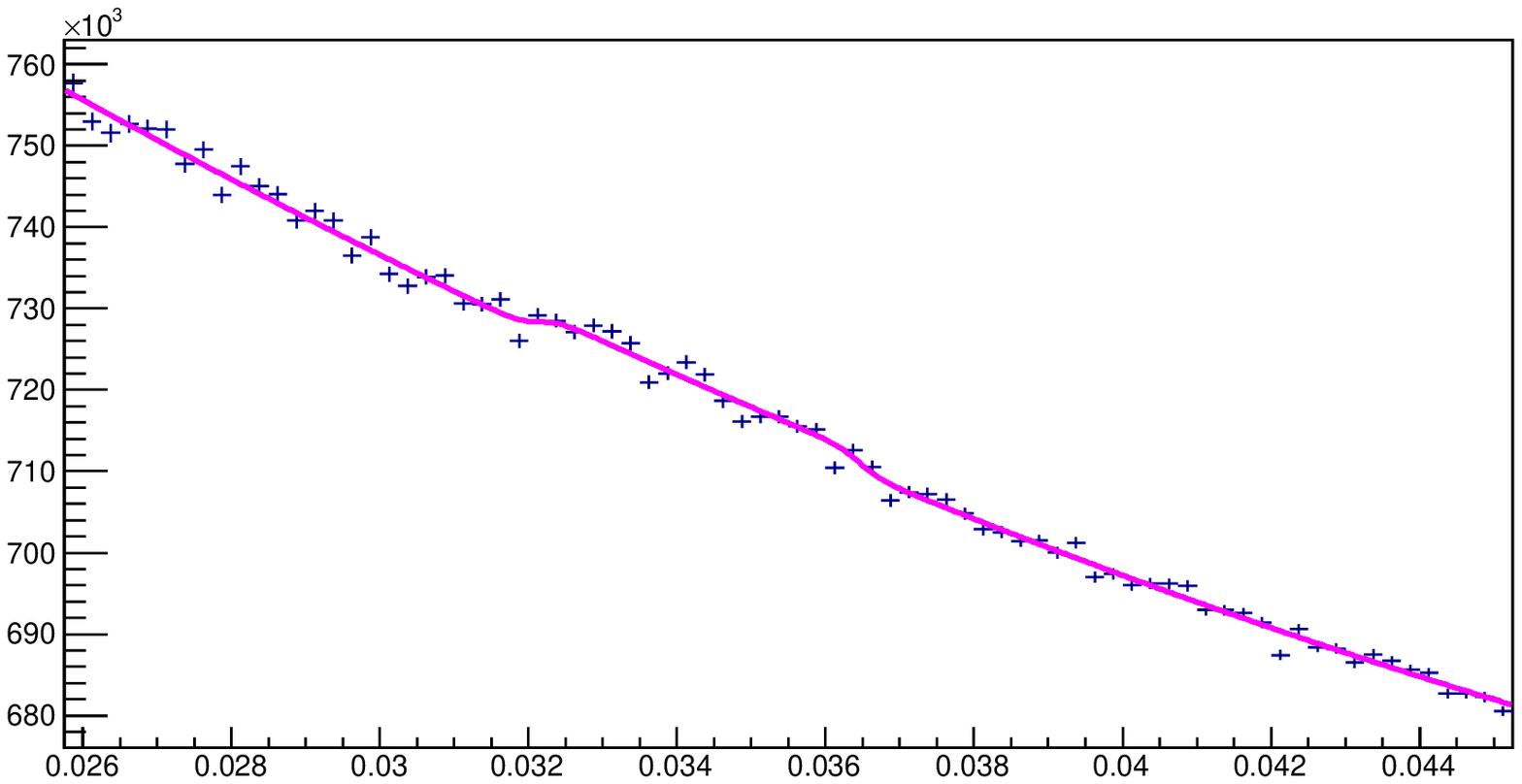}}
\caption{
Fits to histograms of G-LG distances $d$, with $z$ in the range
0.25-0.45 for the northern galactic cap (N), that obtain
$\hat{d}_\alpha$ at $z = 0.36$.
From top to bottom,
they correspond to SDSS DR14 (this analysis), DR14 with galaxy selections
of \cite{BH_ijaa}, and DR13 with galaxy selections
of \cite{BH_ijaa}.
The fits obtain
$\hat{d}_\alpha = 0.03455 \pm 0.00010, 0.03416 \pm 0.00010, 
\textrm{ and } 0.03431 \pm 0.00012$
respectively, where uncertainties are statistical from the fits.
Note that our assigned total uncertainty for $\hat{d}_\alpha$
is $\pm 0.00030$. This single fit for the current analysis,
together with $\theta_*$ obtains
$\Omega_m = 0.272 \pm 0.007$ and $d_* = 0.0345 \pm 0.0004$,
with zero degrees of freedom. 
The relative amplitudes $A$ of the fitted signals are
$0.00552 \pm 0.00060, 0.00369 \pm 0.00042,
\textrm{ and } 0.00341 \pm 0.00039$
respectively. The number of galaxies (G) and
large galaxies (LG) are $(114597, 65130)$, $(153783, 101504)$,
and $(160943, 107971)$, respectively.
Note that the relative amplitude is larger
for the current galaxy selections.
}
\label{fig_250_450_DR13_DR14}
\end{center}
\end{figure}

\section{Comparison with Reference \cite{BH_ijaa}}

Tables 4 and 5 of Reference \cite{BH_ijaa} can be
compared with Tables \ref{galaxy_fits} and \ref{galaxy_theta_fits} of the
present analysis. We find agreement between all measurements
when $d$ in Reference \cite{BH_ijaa} is identified with
$d_*$ in the present analysis.
We find that $d$ in Table 4 of Reference \cite{BH_ijaa}
is biased low with respect to $d_\textrm{\small{drag}}$
in Table \ref{galaxy_fits} of the present analysis.
For the scenario
of flat space and a cosmological constant, Table 4 of Reference \cite{BH_ijaa}
obtains $\Omega_m = 0.284 \pm 0.014$ and $d = 0.0339 \pm 0.0002$.
From this $\Omega_m$ and Equation (\ref{dstar}) we obtain $d_* = 0.0338 \pm 0.0007$,
in good agreement with $d$, so in Reference \cite{BH_ijaa}
no correction for $d_\textrm{\small{drag}}/d_*$ was needed or applied.

\section{Bias of BAO measurements of small galaxy samples}
\label{bias}

We have investigated the difference of $d_\textrm{\small{drag}}$
between Reference \cite{BH_ijaa} and the present analysis.
This difference is not due to the change of data set from 
SDSS DR13 to SDSS DR14: we have compared the coordinates
of selected galaxies and have found no changes in calibrations.
The fluctuation is not caused by the tighter 
galaxy selection requirements of the present analysis: compare 
the entries with ``\&" and ``$*$" in Table \ref{d_meas}, and 
see Figure \ref{fig_250_450_DR13_DR14}. 

As a test,
we divide the bin $0.425 < z < 0.725$ into 6 sub-samples:
$0.425 < z < 0.525$ N, $0.525 < z < 0.625$ N, $0.625 < z < 0.725$ N,
$0.425 < z < 0.525$ S, $0.525 < z < 0.625$ S, and $0.625 < z < 0.725$ S.
We try to fit each one, and average the successful fits 
(only about half are successful),
and obtain $\hat{d}_\alpha = 0.03358 \pm 0.00015$, $\hat{d}_/ = 0.03415 \pm 0.00027$, 
and $\hat{d}_z = 0.03335 \pm 0.00033$.
We also fit the sum of these six bins, and obtain 
$\hat{d}_\alpha = 0.03496 \pm 0.00015$, $\hat{d}_/ = 0.03459 \pm 0.00010$, 
and $\hat{d}_z = 0.03464 \pm 0.00034$.
So there is evidence that fits become biased low as the number of
galaxies is reduced and the significance of the fitted relative amplitude $A$
of the BAO signal becomes marginal. The reason is that the observed
BAO signal has a sharper and larger lower edge at
$\approx 0.032$ compared to the upper edge at $\approx 0.037$,
so the upper edge tends to get lost in the background
fluctuations as the number of galaxies is reduced.


To reduce this bias, in the present analysis we require the
significance of the fitted relative amplitudes $A / \sigma_A > 2$,
instead of $> 1$ for Reference \cite{BH_ijaa}. The price to
pay is that we obtain only 2 independent bins of $z$, instead of 6.

\section{A study of the BAO signal}

The BAO signal has a ``step-up-step-down" shape
with center at $\hat{d}$ and half-width $\Delta$.
The widths of fits vary typically from $\Delta = 0.0017$ to 0.0025,
see Table \ref{selected}.
We have used the center $\hat{d}$ as the  
BAO standard ruler, but could have used the
lower edge of the signal at $\hat{d} - \Delta$,
or the upper edge at $\hat{d} + \Delta$, or
somewhere in between, i.e. $\hat{d} + \epsilon \Delta$.
We have investigated the value of $\epsilon$ that 
minimizes the root-mean-square fluctuations of
a representative selection of 
measurements. 
The result is $\epsilon = -0.17$,
and the difference in the r.m.s. values is
negligible (0.00037 vs. 0.00039) 
so we keep the center of the signal
as our standard ruler, i.e. $\epsilon = 0$.
The r.m.s. fluctuation of the lower edge with $\epsilon = -1$
is 0.00068, and the fluctuation of the upper edge with $\epsilon = 1$
is 0.00091, which again illustrates the bias described
in Appendix \ref{bias}, i.e. the lower edge fluctuates less than
the upper edge. 

A separate open question is whether this center $\hat{d}$
coincides with the $d_\textrm{\small{drag}}$ of
Equation (\ref{ddrag_dstar})?

Yet another question is this: what value of
$\epsilon$ would reproduce the Planck $\Omega_m$?
We obtain $\epsilon$ ranging from $-0.81$ for $\hat{d}_\alpha$
at $z =	0.34$, to $\epsilon = -0.43$ for $\hat{d}_z$ at $z = 0.56$. These large
values of $|\epsilon|$,	and their strong dependence on $z$ and
galaxy-galaxy orientation, do not seem plausible.

Finally, how well do we understand $d_\textrm{\small{drag}} / d_*$?
The present study takes $z_\textrm{\small{drag}} = 1059.94 \pm 0.30$ and
$d_\textrm{\small{drag}} / d_* = 1.0184 \pm 0.0004$ from the
Planck analysis \cite{Planck}. 
Note the extremely small uncertainty obtained by
the Planck Collaboration.
In comparison, from Eq. (4) of Reference \cite{drag}
we obtain $z_\textrm{\small{drag}} = 1020.82$ and
$d_\textrm{\small{drag}} / d_* = 1.044$.

An estimate of the uncertainties due to the issues discussed
in these appendices is included in Table \ref{uncertainties}.

\end{document}